\documentclass[11pt]{article}
\usepackage{cite}
\usepackage{scalerel}
\usepackage{xcolor}
\usepackage{shuffle}
\usepackage{amsmath,amsfonts,amssymb}
\usepackage{latexsym,epsfig}
\usepackage{dsfont}
\usepackage{hyperref}
\usepackage{extarrows}
\usepackage{comment} 
\usepackage{tikz-cd}
\usepackage{amsthm}

\newtheorem*{thm}{Theorem}
\newtheorem*{prop}{Proposition}
\newtheorem*{cor}{Corollary}

\theoremstyle{definition}
\newtheorem*{defn}{Definition}

\def\hybrid{
        \topmargin -20pt
        \oddsidemargin 0pt
        \headheight 0pt \headsep 0pt
        \textwidth 6.25in 
        \textheight 9.5in 
        \marginparwidth .875in
        \parskip 5pt plus 1pt \jot = 1.5ex}

\hybrid

 \csname
@addtoreset\endcsname{equation}{section}

\def\cC{{\cal C}}
\def\cB{{\cal B}}
\def\cG{{\cal G}}
\def\cJ{{\cal J}}

\def\cO{{\cal O}}
\def\cA{{\cal A}}
\def\cE{{\cal E}}

\def\cP{{\cal P}}

\def\cV{{\cal V}}

\def\cK{{\cal K}}

\def\cH{{\cal H}}

\def\del{\partial}

\def\B{\square}

\def\fm{\mathfrak{m}}

\allowdisplaybreaks

\def\bpm{\begin{pmatrix}}
\def\epm{\end{pmatrix}}

\newcommand{\ket}[1]{\lvert#1\rangle}

\begin{document}

\begin{titlepage}
\rightline{}
\rightline{August  2024}
\rightline{HU-EP-24/25}

\begin{center}
\vskip 1.5cm
{\Large \bf{Vertex operators for  the kinematic algebra of Yang-Mills theory}}
\vskip 1.7cm

{\large\bf {Roberto Bonezzi, Christoph Chiaffrino and Olaf Hohm}}
\vskip 1.6cm

\vskip .1cm

\vskip .2cm

\end{center}

\bigskip\bigskip
\begin{center} 
\textbf{Abstract}

\end{center} 
\begin{quote}

The kinematic algebra of Yang-Mills theory can be understood in the framework of homotopy algebras: 
the $L_{\infty}$ algebra of Yang-Mills theory is the tensor product of the color Lie algebra and a 
kinematic space that carries a $C_{\infty}$ algebra. There are also hidden structures that generalize Batalin-Vilkovisky algebras, which explain color-kinematics duality and the double copy but are  
only partially understood. We show that there is a  representation of the $C_{\infty}$ algebra, 
in terms of vertex operators, 
on the Hilbert space of a first-quantized worldline theory.  
To this end we introduce $A_{\infty}$ morphisms, which define the vertex operators and which inject the $C_{\infty}$  algebra into the strictly associative algebra of operators 
on the Hilbert space. 
We also take first steps  to represent the hidden structures on the same space.

\end{quote} 
\vfill
\setcounter{footnote}{0}
\end{titlepage}

\tableofcontents

\vspace{5mm}

\section{Introduction}

Vertex operators  are central pillars in string theory, where  they facilitate  the insertion of string states 
of a scattering process and hence are instrumental for the computation of string scattering amplitudes. 
As such, vertex operators are defined in terms of data of the two-dimensional world-sheet conformal field theory. 
From the perspective of the target space theory they encode on-shell and gauge fixed states. 
More recently, the notion of vertex operators has been suitably broadened so that they can be employed for 
the analysis of off-shell gauge invariant target space theories, including string field theory \cite{Sen:2024nfd}, but also conventional 
gauge theory and gravity \cite{Dai:2008bh,Bonezzi:2018box,Bonezzi:2020jjq}.
In this paper we will further explore such relationships from the viewpoint of the homotopy algebra formulation 
of field theories, such as those of $A_{\infty}$ (or homotopy associative) algebras. In this we were greatly inspired 
by the work of Zeitlin \cite{Zeitlin:2007vd,Zeitlin:2007vv,Zeitlin:2009tj,Zeitlin:2014xma} and previous work by Lian and Zuckerman \cite{Lian:1992mn}.  

Concretely, our work was motivated by the problem to understand the `hidden' kinematic algebra of gauge theories 
such as Yang-Mills theory, which in turn was motivated by the color-kinematics duality 
of scattering amplitudes \cite{Bern:2008qj,Bern:2010ue}. At the level of homotopy algebras this kinematic algebra arises as follows: 
One first uses that any field theory may be encoded in a cyclic $L_{\infty}$ (or homotopy Lie) algebra \cite{Hohm:2017pnh,Jurco:2018sby}. 
For simplicity we will here focus on the case that this  $L_{\infty}$ algebra in turn is derived, by graded symmetrization,
from a simpler $A_{\infty}$ algebra 
with multilinear maps $m_1, m_2, m_3, \ldots$, in which case the action for fields generically denoted by $\phi$ reads 
 \begin{equation} 
  S[\phi] = \frac{1}{2}\langle \phi, m_1(\phi)\rangle + \frac{1}{3} \langle \phi, m_2(\phi,\phi)\rangle +\frac{1}{4} 
  \langle \phi, m_3(\phi,\phi,\phi)\rangle +\cdots \;, 
 \end{equation}
where $\langle\, , \, \rangle$ denotes an inner product on the $A_{\infty}$ algebra. Special cases 
are 3D Chern-Simons theory, where the $A_{\infty}$ algebra reduces to a `strict' differential graded 
associative algebra where all $m_n$ for $n\geq 3$ vanish, and Yang-Mills theory in generic dimensions, 
where in the standard formulation with at most quartic vertices  all $m_n$ for $n\geq 4$ vanish. 
One next uses that the vector space $A$ of the  $A_{\infty}$ or $L_{\infty}$ algebra can be viewed 
as the tensor product of the Lie algebra $\mathfrak{g}$ of the gauge group and a `kinematic' space ${\cal K}$,   
i.e., $A={\cal K}\otimes \mathfrak{g}$, where ${\cal K}$ carries the structure of a homotopy commutative or $C_{\infty}$ algebra \cite{Zeitlin:2008cc,Zeitlin:2009tj}.

Importantly, in addition to this $C_{\infty}$ algebra structure, the kinematic space ${\cal K}$ also carries a vast hidden 
algebraic structure  that is a generalization of a Batalin-Vilkovisky (BV) algebra first identified 
by Reiterer \cite{Reiterer:2019dys},\footnote{This BV algebra should not be confused with the strict BV algebra that emerges when 
formulating Yang-Mills theory in the BV formalism.}  which in turn gives rise to an infinite tower of new 
homotopy maps of ever increasing arity. 
This hidden BV-type algebra on ${\cal K}$ is almost certainly the proper structure that captures 
color-kinematics duality \cite{Ben-Shahar:2021zww,Borsten:2022vtg,Bonezzi:2022bse,Borsten:2023reb,Bonezzi:2023pox} (and that hence enables the `double copy' construction of gravity from gauge theory \cite{Bonezzi:2022yuh,Bonezzi:2022bse,Diaz-Jaramillo:2021wtl,Bonezzi:2023ced,Bonezzi:2023lkx,Borsten:2023ned,Borsten:2023paw,Bonezzi:2024dlv}), 
but for Yang-Mills theory in any local and covariant formulations in generic dimensions the relations of this algebra 
have  so far only been established 
to the order of tri-linear maps, corresponding to the quartic couplings of a field theory. 
It remains as an open problem to find an efficient formulation that would allow one 
to derive the higher homotopy maps from something much simpler, in order to establish all algebraic relations and to 
prove color-kinematics duality directly.

In this paper we explore a strategy using vertex operators 
for deriving the kinematic algebra from something simpler, which  can be 
motivated from the familiar  representation theory of Lie groups.  
Most classical Lie groups one first encounters through their defining or `fundamental'  representation, i.e., 
via explicit operators or matrices acting on a vector space of the smallest possible dimension, 
leaving some natural structure (like a metric) invariant. In contrast, the adjoint representation acts  on 
the Lie algebra itself, which is almost always a space of a larger dimension. 
For instance,  the  exceptional Lie group $E_7$ is 133-dimensional but 
its fundamental representation is of dimension $56$. Clearly, it is computationally cheaper to work with 
$56\times 56$ matrices than to work with $133\times 133$ matrices. (The group $E_8$ is an outlier: its  
$248$-dimensional adjoint representation is the smallest non-trivial representation.) 
All homotopy algebras considered here are infinite-dimensional, but one
may still wonder whether there are `representations' of these homotopy algebras that are easier to 
work with than the underlying algebra itself. We will give an explicit realization of operators acting 
on a vector space, which is the Hilbert space of a first-quantized worldline theory, which contains the 
kinematic  algebra ${\cal K}$ as a subspace. The algebra of operators on the Hilbert space
is necessarily strictly associative and hence much simpler, although the price to pay is that 
the underlying space is vastly bigger.

In order to establish the relation between the strict algebra of operators on the `large' Hilbert space 
and the homotopy algebras on the subspace ${\cal K}$ let us first revisit the notion of a representation of a strict Lie 
algebra $\mathfrak{g}$.  A representation is a linear map 
${\cal V}: \mathfrak{g} \rightarrow {\rm End}(X)$ into the vector space of operators or linear maps (endomorphisms) on a vector space 
$X$ subject to the compatibility condition $[{\cal V}(a), {\cal V}(b)] = {\cal V}([a,b])$. 
Note that on the right-hand side $[\cdot, \cdot]$ denotes the Lie bracket of $\mathfrak{g}$ while on the left-hand side 
it denotes the commutator of operators on $X$, which equips  ${\rm End}(X)$ with a Lie algebra structure. 
Putting this in more sophisticated language: a representation is a morphism between the Lie algebras $\mathfrak{g}$ 
and ${\rm End}(X)$. The representation space $X$ is also called  a module as there is 
an action of $a\in \mathfrak{g}$ on $x\in X$ via $ax:={\cal V}(a)(x)$.

This discussion suggests to define  a representation of an $A_{\infty}$-algebra $(A,\{ m_k\}_{k\geq 1})$ as an 
$A_{\infty}$ morphism from $A$ into $({\rm End}(X),[Q,-],\circ)$, where we indicated  that $X$ now carries 
also a differential $Q$ satisfying $Q^2=0$, which acts on endomorphisms or operators via the commutator, 
hence making $X$ and ${\rm End}(X)$ into  differential graded (dg) vector spaces. 
Moreover, $\circ$ denotes the composition of endomorphisms,  with respect to which $({\rm End}(X),[Q,-],\circ)$ 
becomes a strict $A_{\infty}$-algebra, so that it makes sense to ask whether there is an
$A_{\infty}$ morphism 
$A\rightarrow {\rm End}(X)$. 
As we will explain, in addition to a linear map ${\cal V}_1: A\rightarrow {\rm End}(X)$, 
an $A_{\infty}$ morphism is encoded  in a whole series of multilinear maps 
 \begin{equation}
  {\cal V}_k : A^{\otimes k} \rightarrow {\rm End}(X)\,, \quad k=1,2,3,\ldots\;, 
 \end{equation}
which will play the role of  vertex operators. There is now an infinite number of compatibility conditions
 between the algebraic structures on $A$ and $X$ beginning with: 
  \begin{equation}
  \begin{split}
   [Q, {\cal V}_1(a)] &= {\cal V}_1(m_1(a))\;, \\
   {\cal V}_1(m_2(a,b)) &= {\cal V}_1(a){\cal V}_1(b) + [Q, {\cal V}_2(a,b)] 
   +{\cal V}_2(m_1(a), b) + (-1)^a{\cal V}_2(a, m_1(b)) \;, 
  \end{split}  
  \end{equation} 
where here and in the following we often leave out $\circ$ when the composition of 
endomorphisms is understood. The second equation implies that ${\cal V}_2$ encodes 
the failure of ${\cal V}_1$ to be a conventional algebra morphism.

Using the framework of $A_{\infty}$ morphisms we will show explicitly how to inject the kinematic 
$C_{\infty}$ algebra of Yang-Mills theory, which is a genuine homotopy algebra, into the strictly 
associative algebra of operators on a Hilbert space, thereby relating the former to the latter. However, as we will explain,  this map is not 
a so-called quasi-isomorphism, 
as would be the case if this mapping was homotopy 
transfer.\footnote{See \cite{Kajiura:2003ax,Nutzi:2018vkl,Arvanitakis:2019ald,Arvanitakis:2020rrk,Chiaffrino:2020akd,Arvanitakis:2021ecw,Okawa:2022sjf,Bonezzi:2023xhn,Borsten:2024cfx} for field theory applications of homotopy transfer.} (This mapping 
should be a quasi-isomorphism if we wanted to claim that the homotopy algebra is really derived from the strict algebra.) 
Even though the relation we establish is still a conceptually satisfying simplification it must be admitted that, in a sense, 
the complexity of the algebra has been moved to the vertex operators or 
$A_{\infty}$ morphisms, which for Yang-Mills theory 
feature a non-trivial ${\cal V}_2$ but no higher maps. 
We will also show, however, that the vertex operators themselves define a dg associative algebra, 
which we term `vertex algebra'\footnote{This should not be confused with the concept of vertex operator algebras.} 
and which consist of a special class of so-called `twisting morphisms'. 
The vertex operators or $A_{\infty}$ morphisms, collectively called ${\cal V}$,  are then 
those that satisfy the corresponding Maurer-Cartan equation of this dg algebra: 
$\partial \mathcal{V} + \mathcal{V} \circ \mathcal{V} = 0$.

The rest of this paper is organized as follows: In sec.~2 we discuss the mathematical machinery of $A_{\infty}$ algebras 
and their morphisms and introduce the above mentioned vertex algebra. 
We also argue that the deformations of a BRST-type operator $Q$ are governed by such vertex operators. 
Even though we recall basic notions 
to establish our conventions this section is not completely self-contained in that we assume familiarity with homotopy algebras, 
as for instance reviewed in \cite{Hohm:2017pnh,Arvanitakis:2020rrk,Kajiura:2003ax,Bonezzi:2023xhn}. In sec.~3 we turn to our core example of the kinematic algebra of Yang-Mills theory. 
We introduce the Hilbert space of a world-line theory that was recently explored by one of us \cite{Bonezzi:2024emt}, 
and we determine the Vertex operators mapping the kinematic algebra of Yang-Mills to the associative algebra 
of operators on the Hilbert space. Secs.~2 und 3 can largely be read independently, and readers not so comfortable 
with homotopy algebras are advised to read through the example of sec.~3 first. 
In sec.~4 we begin exploring the hidden BV-type algebra that we are most interested in, as this is required 
for understanding color-kinematics duality and the double copy. These results are still tentative, but we believe that 
setting up this framework is a promising step in understanding the hidden kinematic algebra of Yang-Mills theory. 
We close with our conclusions in sec.~5. In the appendix we give a general discussion of how to `strictify' 
homotopy algebras, i.e., of how to render them strict by passing over to a larger vector space. While the particular strictification we present is 
taken from \cite{Amorim_2016}, we add a number of details and explanations that hopefully make it more accessible to non-experts.

\section{Vertex operators and $A_\infty$ modules}\label{sec:modules}

\subsection{Generalities}

This work strongly relies on the theory of homotopy associative algebras, also known as $A_\infty$ algebras. Below we recall some important definitions and facts and also establish our notation. The reader unfamiliar with these concepts can find more details for example in \cite{Kajiura:2003ax,Arvanitakis:2020rrk,Bonezzi:2023xhn} .

\paragraph*{Chain complexes:} A chain complex $(X,Q)$ is a graded vector space $X$ equipped with a degree one map $Q$  satisfying $Q^2 = 0$. We write $X^i \subseteq X$ for the subspace of elements of degree $i$. The homology of $(X,Q)$ is the quotient space  $H(X) = \frac{\ker Q}{\text{im} \, Q}$. Given chain complexes $(X,Q_1)$ and $(Y,Q_2)$, a chain map is a degree zero linear map $f: X \rightarrow Y$, such that
\begin{equation}
f \circ Q_1 = Q_2 \circ f \, .
\end{equation}
This property ensures that $f$ induces a linear map $H(f): H(X) \rightarrow H(Y)$ on homology. A chain map $f$ is called a quasi-isomorphism if $H(f)$ is an isomorphism, i.e.,  if it is invertible.

\paragraph*{Suspension:} Let $X$ be a graded vector space. We define the suspension $sX$, such that
\begin{equation}
(sX)^i = X^{i+1} \, .
\end{equation}
Given a multilinear map of graded vector spaces
\begin{equation}
f: X_1 \otimes \cdots \otimes X_n \longrightarrow Y \, ,
\end{equation}
we define the suspended map $\text{dec}(f)$
\begin{equation}
\begin{split}
\text{dec}(f): sX_1 \otimes \cdots \otimes sX_n &\longrightarrow sY \, , \\
x_1 \otimes \cdots \otimes x_n &\longmapsto (-)^{nf + \sum_{i = 1}^{n}(n-i)x_i} f(x_1,\ldots, x_n) \, ,
\end{split}
\end{equation}
where on the right hand side the degrees in the exponent are taken with respect to $X$ and $Y$. We note that $\text{dec}(f)$ has degree $|\text{dec}(f)| = |f| + n - 1$.   For details on the origin of the signs, see for example section 1 of \cite{FiorenzaDec}. Henceforth, we will often write $f$ instead of $\text{dec}(f)$ whenever it is clear from the context whether $f$ is considered as a map on the spaces $X$ or their suspension.

\paragraph*{$A_\infty$ algebras:} An $A_\infty$ algebra (or homotopy associative algebra) $(A, \{m_n\}_{n \ge 1})$ is a chain complex $(A,m_1)$ together with a collection of multilinear maps
\begin{equation}
m_n: A^{\otimes n} \longrightarrow A
\end{equation}
of degree $2-n$ satisfying certain relations. We often refer to the $m_n$ as products. A morphism $f: (A, \{m_n^A\}_{n \ge 1}) \rightarrow (B, \{m_n^B\}_{n \ge 1})$ consists of a collection of multilinear maps
\begin{equation}
f_n: A^{\otimes n} \longrightarrow B\;, \quad n\geq 1\;, 
\end{equation}
of degree $1-n$ satisfying relations with respect to the products $\{m_n^A\}_{n \ge 1}$ and $\{m_n^B\}_{n \ge 1}$. The first of these relations is that $f_1: (A,m_1^B) \rightarrow (B,m_1^B)$ is a chain map between chain complexes. We say that $f$ is a quasi-isomorphism if $f_1$ is. We will usually denote the collection $\{f_n\}_{n \ge 1}$ by a single letter $f$.

\paragraph*{Maurer-Cartan equation:} Let $(A,\{m_n\}_{n \ge 1})$ be an $A_\infty$ algebra. Given an element $a \in A$ of degree one, we define the Maurer-Cartan equation
\begin{equation}
\text{MC}(a) := m_1(a) + m_2(a,a) + m_3(a,a,a) + \ldots = 0 \, .
\end{equation}
Elements $a \in A$ satisfying the Maurer-Cartan equation are called Maurer-Cartan elements.

\paragraph*{Differential graded coalgebras:} A differential graded coalgebra $(C,M_C,\Delta_C)$ is a differential graded vector space $(C,M_C)$, together with a degree zero linear map
\begin{equation}
    \Delta_C: C \longrightarrow C \otimes C
\end{equation}
called the coproduct. $M_C$ has to be a coderivation, which means that it satisfies
\begin{equation}
    \Delta \circ M_C = (M_C \otimes \text{id}_C + \text{id}_C \otimes M_C) \circ \Delta_C \, .
\end{equation}
Further, $\Delta$ is required to be coassociative, i.e.
\begin{equation}
    (\Delta \otimes \text{id}_C) \circ \Delta =  (\text{id}_C \otimes \Delta) \circ \Delta \, .
\end{equation}
A morphism of differential graded coalgebras is a linear map $
F: (C,M_C,\Delta_C) \rightarrow (D,M_D,\Delta_D)$, such that
\begin{equation}
    \Delta_D \circ F = (F \otimes F) \circ \Delta_C \, .
\end{equation}
A differential graded coalgebra with zero differential is simply called a graded coalgebra. 

\paragraph{The tensor coalgebra:} Given a graded vector space $V$, we can construct the tensor coalgebra $(T^c(V),\Delta)$, where $T^c(V) = \bigoplus_{n \ge 0} V^{\otimes n}$ and
\begin{equation}
\Delta(v_1 \cdots v_n) = 1 \otimes v_1 \cdots v_n + v_1 \cdots v_n \otimes 1 + \sum_{i = 1}^{n-1} v_1 \cdots v_{i} \otimes v_{i+1} \cdots v_n \, .
\end{equation}
Here, $v_1 \cdots v_n$ denotes an element in $V^{\otimes n}$, so that the right hand side is an element of $T^c(V) \otimes T^c(V)$. Any coderivation $M$ on $T^c(V)$ is fully determined by a linear map
\begin{equation}
    m: T^c(V) \longrightarrow V \, ,
\end{equation}
so that $\pi_1 \circ M = m$, with $\pi_1: T^c(V) \rightarrow V$  the projection to $V \subseteq T^c(V)$. 
For instance, $M$ could be a degree-1 coderivation so that $(T^c(V),M, \Delta)$ is a differential graded coalgebra. 
Similarly, a morphism $F: (C,\Delta_C) \rightarrow (T^c(V),\Delta)$ from any graded coalgebra $C$ into $T^c(V)$ is fully determined by a degree zero linear map
\begin{equation}
    f: T^c(V) \longrightarrow V \, ,
\end{equation}
so that $\pi_1 \circ F = f$.

The relations and morphisms of $A_\infty$ algebras can be conveniently described in the language of differential graded coalgebras. 
In particular, the $A_\infty$  relations are encodes in $M^2=0$ for a degree-1 
coderivation $M$.  
This is known as the bar construction.
\begin{thm}[Bar construction]
Let $f: (A, \{m_n^A\}_{n \ge 1}) \rightarrow (B, \{m_n^B\}_{n \ge 1})$ be a morphism of $A_\infty$ algebras. Then $f$ can be equivalently described as a morphism $\mathcal{B}(f)$ of differential graded coalgebras
\begin{equation}
\mathcal{B}(f): (\mathcal{B}(A),M_A,\Delta) \longrightarrow (\mathcal{B}(B),M_B,\Delta) \, ,
\end{equation}
where $\mathcal{B}(A) = T^c(sA)$ is the tensor coalgebra over the suspended vector space $A$ with coproduct $\Delta$, and $M_{A}$, resp.~$M_B$, are coderivations defined in terms of $\{m^{A}_n\}_{n \ge 1}$, resp.~$\{m^{B}_n\}_{n \ge 1}$. Explicitly, $\mathcal{B}(f)$, $M_A$, $M_B$ are such that
\begin{equation}
\pi_1 \circ \mathcal{B}(f) = \sum_{i \ge 1} \mathrm{dec}(f_i) \ , \quad \pi_1 \circ M_{A/B} = \sum_{i \ge 1} \mathrm{dec}(m^{A/B}_{i}) \, .
\end{equation}
Note that, since $|f_i| = 1 - n$, we have that $|\mathrm{dec}(f_i)| = 0$ independent of $i$. Similarly, we have $|\mathrm{dec}(m_i)| = 1$.
\end{thm}

Differential graded algebras are both an associative algebra and a chain complex, so that the differential of the chain complex satisfies the Leibniz rule. $A_\infty$ algebras are a generalization of differential graded algebras.  

\paragraph{Differential graded algebras:}
A differential graded algebra $(A,m_1,m_2)$ is an $A_\infty$ algebra $(A,\{m_n\}_{n \ge 1})$, such that all $m_n = 0$ for $n \ge 3$. Differential graded algebras are also called strict $A_\infty$ algebras.

\subsection{$A_{\infty}$-modules}

\begin{defn}[Algebra of endomorphisms]
Given a chain complex $(X,Q)$, we can associate to it the differential graded algebra $(\text{End}(X),[Q,-],\circ)$, where $\text{End}(X)$ is the graded vector space of linear maps $V: X \rightarrow X$ which has the composition product $(V,W) \mapsto V \circ W$. The differential is defined with respect to the commutator bracket
\begin{equation}
[V,W] := V \circ W - (-)^{VW} W \circ V \, ,
\end{equation}
so that $[Q,-]$ acts on $V$ as
\begin{equation}
[Q,V] = Q \circ V - (-)^V V \circ Q \, .
\end{equation}
We call $({\rm End}(X),[Q,-],\circ)$ the algebra of endomorphisms on $X$.
\end{defn}
The endomorphism algebra can be used to generalize the concept of modules to apply also to $A_\infty$ algebras. We recall that, given an associative algebra $(A,m_2)$, an $A$-module $X$ is a vector space, on which $A$ acts from the left. In other words, there is a map
\begin{equation}\label{AModule1}
\mathcal{V}_1^\dagger: A \otimes X \longrightarrow X \, ,
\end{equation}
such that 
\begin{equation}\label{Amoduleprop1}
\mathcal{V}_1^\dagger(a,\mathcal{V}^\dagger_1(b,x)) = \mathcal{V}^\dagger_1(m_2(a,b),x))
\end{equation}
for all $a,b \in A$ and $x \in X$. One often writes $\mathcal{V}^\dagger_1(a,x) := ax$. Thus, an $A$-module is very similar to a vector space, but defined over the algebra $A$ rather than only over the underlying field. In particular, we can take superpositions of elements in $X$ over elements in $A$. For example, for $a,b \in A$ and $x,y \in X$, the combination
\begin{equation}
ax + by := \mathcal{V}^\dagger_1(a,x) + \mathcal{V}_1^\dagger(b,y) 
\end{equation}
exists as an element in $X$.

An equivalent characterization of an $A$-module is the following. Note that the map $\mathcal{V}_1^\dagger$ in \eqref{AModule1} can be equivalently described as a map
\begin{equation}\label{AModule2}
\mathcal{V}_1: A \longrightarrow \text{End}(X) \, ,
\end{equation}
where an element $a \in A$ is mapped to the endomorphism $x \mapsto \mathcal{V}^\dagger_1(a,x)$. Since  $\mathcal{V}_1$ in \eqref{AModule2} is a map between associative algebras we can require that $\mathcal{V}_1$ is a morphism of algebras, meaning that
\begin{equation}
\mathcal{V}_1(m_2(a,b)) = \mathcal V_1(a) \circ \mathcal V_1(b) \, .
\end{equation}
It is easy to check that this condition is equivalent to $\mathcal{V}^\dagger_1$ satisfying \eqref{Amoduleprop1}. Therefore, an $A$-module structure $\mathcal{V}_1^\dagger$ \emph{is} a morphism of associative algebras $\mathcal{V}_1$. This observation motivates the following definition.\footnote{This definition can be found for example in \cite{keller2001}, although there the author considers the elements of $A$ to act on $X$ from the right instead of them acting from the left. In the appendix, a treatment of both left and right module structures can be found.}
\begin{defn}
Let $(A,\{m_n\}_{n \ge 1})$ be an $A_\infty$ algebra. An $A$-module $X$ is a differential graded vector space $(X,Q)$, together with a morphism of $A_\infty$ algebras
\begin{equation}\label{ModuleDef}
\mathcal{V}: (A,\{m_n\}_{n \ge 1}) \longrightarrow (\text{End}(X),[Q,-],\circ) \, .
\end{equation}
\end{defn}

Recall that $\mathcal{V}$ being a morphism means that it is defined in terms of a collection of multilinear maps
\begin{equation}
\mathcal{V}_k: A^{\otimes k} \longrightarrow X
\end{equation}
 of degree $1-k$. In the special case of $A$-modules of associative algebras $A$, $\mathcal{V}_1$ is just the linear part.

As an example, we note that any $A_\infty$-algebra is a module over itself. This generalizes the fact that, given an associative algebra $(A,m_2)$, we can define
\begin{equation}
\mathcal{V}_1(a) = m_2(a,-) \, .
\end{equation}
For a genuine $A_\infty$ algebra $(A, \{m_n\}_{n \ge 1})$, this can be generalized by defining
\begin{equation}\label{Aoveritself}
\mathcal{V}_n(a_1,\ldots,a_n) = m_{n+1}(a_1,\ldots,a_{n},-) \, ,
\end{equation}
see for example \cite{keller2006}, in particular section 3 therein. In the appendix, we explain how the $A_\infty$ module structure on $A$ itself can be used to find a differential graded algebra $\text{End}^\infty_A(A)$ quasi-isomorphic to $A$. In this case, one says that $\text{End}^\infty_A(A)$ is a strictification of $A$.\footnote{In contrast to the construction given in the appendix, although any $A$-module is given by an $A_\infty$ morphism into $(\text{End}(X),[Q,-],\circ)$, it will almost never be a strictifictation of $(A,\{m_n\}_{n \ge 1})$, for  this would imply that $H(\text{End}(X)) = \text{End}(H(X))$ is isomorphic to $H(A)$, which is certainly almost never true when $A = X$.}

\subsection{Characterization of $A_\infty$ modules as twisting morphisms}

$A_\infty$ algebra morphisms can themselves be embedded into a differential graded algebra. This algebra is defined in the following theorem (see section 1.6. and \textbf{Proposition 2.1.2.} of \cite{loday2012algebraic}): 
\begin{thm}
Let $(C,\text d_C, \Delta)$ a differential graded coalgebra and $(A,\text d_A, \mu)$ be a differential graded algebra. We associate to it the \textbf{convolution algebra} $(\hom(C,A),\partial, \star)$, where $\hom(C,A)$ is the space of linear maps $f: C \rightarrow A$, the differential $\partial$ is
\begin{equation}
\partial f = \text d_A \circ f - (-)^f f \circ \text d_C
\end{equation}
and the product is given by
\begin{equation}
f \star g = \mu \circ (f \otimes g)\circ \Delta \, .
\end{equation}
Then 
$(\hom(C,A),\partial, \star)$ defines a differential graded algebra. Degree one elements $f$, which solve the Maurer-Cartan equation
\begin{equation}\label{twMC}
0 = \partial f + f \star f \,, 
\end{equation}
are called \textbf{twisting morphisms}.
\end{thm}
The definition of the product $\star$ can be viewed as a generalization of the product arising as the dual of a coalgebra. Given any coalgebra $(C,\Delta)$, its linear dual $C^* = \hom(C,\mathbb{R})$ is an algebra. In the above definition, this is the case $A = \mathbb{R}$.

Suppose we have an $A_\infty$ morphism $f: (A,\{m_n^A\}_{n \ge 1}) \rightarrow (B,m_1^B,m_2^B)$ from an $A_\infty$ algebra to a differential graded algebra. Recall that the bar construction associates to it a map of differential graded coalgebras
\begin{equation}
\mathcal B(f): (\mathcal{B}(A),M_A,\Delta) \longrightarrow (\mathcal{B}(B),M_B,\Delta) \, ,
\end{equation}
which can also be described as a degree one linear map
\begin{equation}
F: \mathcal{B}(A) \longrightarrow B
\end{equation} 
with $F = \pi_1 \circ \mathcal{B}(f)$.\footnote{Since $F = \pi_1 \circ \mathcal{B}(f)$ has degree zero as a map $F: \mathcal{B}(A) \rightarrow sB$, it has degree one as a map $F: \mathcal{B}(A) \rightarrow B$.} Since, by assumption, $(B,m_1^B,m_2^B)$ is a differential graded algebra, the map $F$ is a linear map from a differential graded coalgebra $(\mathcal{B}(A),M_A,\Delta)$ into a differential graded algebra $(B,m_1^B,m_2^B)$. We state  the following fact, which follows as a special case of \textbf{Theorem 2.2.9.} in \cite{loday2012algebraic}: 
\begin{thm}
$f$ is an $A_\infty$ morphism if and only if $F = \pi_1 \circ \mathcal{B}(f)$ satisfies the Maurer-Cartan equation \eqref{twMC}.
\end{thm}
We can apply this theorem to characterize modules of $A_\infty$ algebras. An $A$-module $(X,Q)$ over an $A_\infty$ algebra $(A,\{m_n\}_{n \ge 1})$ is a twisting morphism $\mathcal{V}$ of $(\hom(\mathcal{B}(A),\text{End}(X)),\partial, \star)$.
We provide explicit formulas for later reference. For an element $\mathcal{V} \in \hom(\mathcal{B}(A),\text{End}(X))$, we write
\begin{equation}
\mathcal{V} = \sum_{k \ge 1} \mathcal{V}_k \, ,
\end{equation}
where $\mathcal{V}_k: (sA)^{\otimes k} \rightarrow \text{End}(X)$ is the $k$-multilinear part of $\mathcal{V}$. The differential $\partial$ is given by
\begin{equation}
\begin{split}
(\partial \mathcal V)_{l}(a_1,\ldots,a_l) &= [Q,  \mathcal{V}_l(a_1,\ldots,a_l)] \\ & \quad -  \sum_{j + k = l + 1} \sum_{i = 1}^{ l - j +1} (-)^{\mathcal{V} + a_1 + \ldots + a_{i-1}}\mathcal{V}_k(a_1,\ldots, \mathrm{dec}(m_{j})(a_{i},\ldots),a_{i+j},\ldots) \, ,
\end{split}
\end{equation}
where the degrees of the $a_i$ are with respect to the shifted vector space $sA$.
The product $\star$ on elements $\mathcal{V},\mathcal{W} \in \hom(\mathcal{B}(A),\text{End}(X))$ is
\begin{equation}
(\mathcal{V} \star \mathcal{W})_k(a_1,\ldots,a_k) = \sum_{i+j = k} (-)^{\mathcal{W}_j(a_1 + \ldots + a_i)} \mathcal{V}_i(a_1,\ldots,a_i) \circ \mathcal{W}_j(a_{i+1},\ldots, a_k) \, .
\end{equation}

In the appendix, we will give another description of $A_\infty$ modules in terms of differential graded comodules. This is a variant of the bar construction.

\subsection{Deformations of BRST operator}

We will now argue  that $A_\infty$ modules appear naturally as background dependent deformations of BRST operators.
BRST quantization gives rise to a BRST charge $Q$, which is a differential on a graded vector space $X$, so that $(X,Q)$ defines a chain complex. The elements of $X$ in degree one are considered fields $\psi$ of a physical system, with equation of motion
\begin{equation}\label{QEOM}
Q\psi = 0
\end{equation}
and gauge transformations
\begin{equation}
\delta_\lambda \psi = Q \lambda \, ,
\end{equation}
where $\lambda$ is of degree zero. The equations of motion are gauge independent due to $Q^2 = 0$.

Deformations of $Q$ are perturbations
\begin{equation}\label{ConstantPert}
Q \longmapsto Q_V = Q + V \, ,
\end{equation}
where $V \in \text{End}(X)$ is of degree one. We require that the perturbation preserves the square zero condition, i.e., 
\begin{equation}\label{ConstantPert2}
Q_V^2 = (Q + V)^2 = [Q,V] + V \circ V = 0 \,. \end{equation}
Therefore, the allowed deformations are exactly the Maurer-Cartan elements of the differential graded algebra $(\text{End}(X),[Q,-],\circ)$.

Since the equations of motion \eqref{QEOM} depend only linearly on $\psi$, a chain complex $(X,Q)$ can only describe free theories. To obtain some notion of interaction, we can try to find a description of the field $\psi$ propagating in some background, described by another field $\phi$. Suppose that the theory of the background field $\phi$ can be encoded in an $A_\infty$ algebra $(A,\{m_n\}_{n \ge 1})$, with equations of motion given by the Maurer-Cartan equation
\begin{equation}\label{BackgroundMC}
\text{MC}(\phi) = m_1(\phi) + m_2(\phi,\phi) + \ldots = 0 \, .
\end{equation}

The idea is that in the presence of a background $\phi$, a BRST operator $Q_\phi$ only squares to zero when the background field $\phi$ is on-shell (i.e.~when it satisfies \eqref{BackgroundMC}). One therefore demands that
\begin{equation}\label{OffshellPert}
Q^2_\phi = f_\phi(\text{MC}(\phi)) \, ,
\end{equation}
where the right hand side depends linearly on the Maurer-Cartan equation of $\phi$ (but in general non-linearly on $\phi$).

Given a background independent BRST operator $Q$, one can try to introduce a background dependence perturbatively by making $V$ in \eqref{ConstantPert} field dependent. Therefore, we consider perturbations
\begin{equation}
Q \longmapsto Q_V = Q + V_\phi \, ,
\end{equation}
such that $Q_V$ squares to zero up to the Maurer-Cartan equation \eqref{BackgroundMC}.

Background dependent deformations have been previously considered in terms of vertex operators, see for example \cite{Dai:2008bh,Bonezzi:2018box,Bonezzi:2020jjq}, which all consider deformations of a BRST operator of a given worldline theory. They arrive at a condition of the form \eqref{OffshellPert}, although for cases where the Maurer-Cartan equation is that of an $L_\infty$ algebra. More general deformations were also considered in string field theory, where one can deform the closed string background, in which an open string propagates, see  \cite{Kajiura:2004xu, Kajiura:2005sn, Moeller:2010mh, Munster:2011ij, Munster:2012gy}. In these cases, not only the BRST charge of the open string is deformed, but also the higher vertices.

When dealing with deformations of BRST charges like in worldline theories, we propose that the vertex operators $V_\phi$ are encoded in $A$-module structures $\mathcal{V}$ on $(X,Q)$.\footnote{We want to emphasize that all constructions here apply almost immediately to the case of $L_\infty$ algebras, where a similar Maurer-Cartan equation like \eqref{twMC} exists. Solutions to that equation are then $L_\infty$ representations.} Given such a $\mathcal{V} = \sum_{i \ge 1} \mathcal{V}_i$, we can define 
\begin{equation}\label{TransportedMC}
V_\phi = \sum_{i \ge 1}\mathcal{V}_i(\phi,...,\phi) \, .
\end{equation}
We recall that
\begin{equation}
\mathcal{V}: (A,\{m_n\}_{n \ge 1}) \longrightarrow (\text{End}(X),[Q,-],\circ)
\end{equation}
is an $A_\infty$ morphism, which implies that it preserves Maurer-Cartan elements, meaning that if $\phi$ is a Maurer-Cartan element of $(A,\{m_n\}_{n \ge 1})$, then $V_\phi$ defined in \eqref{TransportedMC} is a Maurer-Cartan element of $(\text{End}(X),[Q,-],\circ)$. But as we saw in \eqref{ConstantPert2}, these are exactly the allowed perturbations of $Q$. Furthermore, the fact that $\mathcal{V}$ satisfies the equation of a twisting morphism \eqref{twMC} implies an equation of the form \eqref{OffshellPert}. In order to prove this, recall that the twisting equation takes the form
\begin{equation}
[Q,\mathcal V] - \mathcal V \circ M_A + \mathcal{V} \star \mathcal{V} = 0\,.
\end{equation}
Now let $\phi \in A$ be an element of degree one. The twisting equation, when evaluated on the formal infinite sum
\begin{equation}
\exp \phi := \sum_{k \ge 0} \phi^{\otimes k}\;, 
\end{equation}
becomes
\begin{equation}\label{MConfields}
[Q,V_\phi] - \mathcal{V}(e^\phi \otimes \text{MC}(\phi) \otimes e^\phi) + V_\phi \circ V_\phi = 0 \, ,
\end{equation}
where
\begin{equation}
\mathcal V(e^\phi \otimes \text{MC}(\phi) \otimes e^\phi) = \sum_{k,l \ge 0}\mathcal{V}_{k+l+1}(\phi^{\otimes k} \otimes \text{MC}(\phi) \otimes \phi^{\otimes l}) \, .
\end{equation}
When moving this expression to the other side of \eqref{MConfields}, we find that
\begin{equation}
[Q,V_\phi]  + V_\phi \circ V_\phi = \mathcal{V}(e^\phi \otimes \text{MC}(\phi) \otimes e^\phi) \, ,
\end{equation}
which is exactly of the form \eqref{OffshellPert}.

As an example, suppose we have an $A_\infty$ algebra $(A,\{m_k\}_{k \ge 1})$. We can think of $(X,Q) = (A,m_1)$ as a BRST complex. The vertex operators given in \eqref{Aoveritself} then define a deformation of $(X,Q)$ over the background described by the $A_\infty$ algebra $(A,\{m_k\}_{k \ge 1})$\footnote{See e.g. \cite{Zeitlin:2008cc} for an explicit example in Yang-Mills theory.}.

\section{Worldline vertex operators and the $C_\infty$ algebra of Yang-Mills}

In this section we will show that the $C_\infty$ algebra of Yang-Mills theory can be generated through the vertex operators of a worldline theory, giving a concrete example of the general formalism described in section \ref{sec:modules}. The vertex operators to be considered are quantum mechanical operators acting on the BRST Hilbert space of the worldline model, which we describe in the following.

\subsection{Worldline Hilbert space and operator algebra}

We begin by defining the worldline Hilbert space and the class of operators acting on it. These arise from the Hamiltonian BRST quantization of the bosonic spinning particle \cite{Henneaux:1987cp,Bouatta:2004kk,Hallowell:2007qk,Bastianelli:2009eh,Cherney:2009mf,Bonezzi:2024emt}. The quantized phase space consists of the bosonic canonical pairs $(x^\mu,p_\nu)$ and $(\alpha^\mu,\bar\alpha^\nu)$ obeying
\begin{equation}\label{commBose}
[x^\mu,p_\nu]=i\,\delta^\mu_\nu\;,\quad[\bar\alpha^\mu,\alpha^\nu]=\eta^{\mu\nu}\;,    
\end{equation}
together with fermionic canonical pairs $(b,c)$, $(\bar \cB,\cC)$ and $(\bar\cC,\cB)$, obeying
\begin{equation}\label{commFermi}
\{b,c\}=1\;,\quad \{\bar\cB,\cC\}=1\;,\quad\{\bar\cC,\cB\}=1\;,   
\end{equation}
with all other (anti)commutators vanishing. In particular, all fermionic operators are nilpotent. $x^\mu$ and $p_\nu$ are the usual position and momentum operators in quantum mechanics, with $\mu=0,\ldots,D-1$ a spacetime Lorentz index. The bosonic oscillators $(\bar\alpha^\mu,\alpha^\nu)$ generate spacetime spin degrees of freedom, while all fermionic operators are BRST ghosts. We define the ghost number operator
\begin{equation}
\cG:=cb+\cC\bar\cB-\cB\bar\cC\;,    
\end{equation}
which assigns ghost degree $+1$ to $(c,\cC,\bar\cC)$ and $-1$ to $(b,\cB,\bar\cB)$, while all bosonic canonical pairs have degree zero.

As Hilbert space we choose the tensor product of smooth functions of $x$, on which $p_\mu$ acts as $-i\del_\mu$, with a Fock space of oscillators. The latter is defined by a vacuum state $\ket{0}$ obeying
\begin{equation}
(b,\bar\cB,\bar\cC,\bar\alpha^\mu)\ket{0}=0\;,    
\end{equation}
and is thus generated by monomials in $(c,\cC,\cB,\alpha^\mu)$. The Hilbert space $\cH$ is graded with respect to the ghost number $\cG$, with eigenvalues from $-1$ to $+2$. One can define a second integer degree, ranging from zero to infinity, as the eigenvalue with respect to the following $U(1)$ charge:
\begin{equation}
\cJ:=\alpha^\mu\bar\alpha_\mu+\cC\bar\cB+\cB\bar\cC=N_\alpha+N_\cC+N_\cB \;,   
\end{equation}
which acts on $\cH$ by counting the total number of $(\cC,\cB,\alpha^\mu)$ modes. Since $\cJ$ and $\cG$ commute with each other, the Hilbert space decomposes as the double direct sum
\begin{equation}
\cH=\bigoplus_{s=0}^\infty\cH_s\;,\quad\cH_s=\bigoplus_{k=-1}^2\cH_{s,k}\;, 
\end{equation}
with respect to eigenspaces $\cH_{s,k}$ with charge $\cJ=s$ and ghost number $\cG=k$.

The BRST charge $Q$ is a nilpotent operator of ghost number $+1$, which we take to be
\begin{equation}\label{Q}
Q=c\,\B+\big(\bar\cC\alpha^\mu+\cC\bar\alpha^\mu\big)\,\del_\mu-\cC\bar\cC \,b\;, 
\end{equation}
where $\B=\del^\mu\del_\mu$ is the wave operator. Importantly, $Q$ is neutral under the $U(1)$ charge, $[\cJ,Q]=0$, implying it acts diagonally on the eigenvalue $s$: 
\begin{equation}
Q:\cH_{s,k}\rightarrow\cH_{s,k+1} \;.   
\end{equation}
Each subspace $(\cH_s,Q)$ is a chain complex and the BRST cohomology describes a tower of free massless particles of spin $s,s-2,\ldots,$ down to spin one or zero \cite{Bouatta:2004kk,Bonezzi:2024emt}. From now on we will restrict to the subspace $\cH_1$, which carries the kinematic algebra of Yang-Mills theory.

General operators acting on $\cH$ are given by tensor products of functions of $x$ with polynomials in $\del_\mu$ and in all creation and annihilation operators. The operators of charge zero, meaning they obey $[\cJ,\cO]=0$, form a subalgebra and can be restricted to endomorphisms of $\cH_1$, i.e.~$\cO:\cH_1\to\cH_1$. Upon collectively denoting $\alpha^M:=(\cC,\alpha^\mu,\cB)$ the creation operators with charge $+1$, a generic operator $\cO$ on $\cH$ of charge zero can be written in normal ordered form  as
\begin{equation}\label{zerocharge}
\cO=\sum_{k=0}^\infty \cO_{(k)}\;,\quad \cO_{(k)}:= O_{M_1\ldots M_k\,N_1\ldots N_k}\,\alpha^{M_1}\cdots \alpha^{M_k}\,\bar\alpha^{N_1}\cdots\bar\alpha^{N_k}\;, 
\end{equation} 
where the operator coefficients $O_{M_1\ldots M_k\,N_1\ldots N_k}$ still depend on $b,c,x^\mu,\del_\mu$. 
A general element of $\cH_1$ takes the form
\begin{equation}
\psi=\psi_M(x,c)\,\alpha^M\ket{0}\;,   
\end{equation}
and is thus annihilated by all $\cO_{(k)}$ with $k>1$, which form an ideal under composition. To see this, consider two operators $\cO=\cO_{(n)}$, with $n>1$, and $\cO'=\cO'_{(0)}+\cO'_{(1)}$. Rewriting the composition in normal ordering we have
\begin{equation}\label{ideal}
\begin{split}
\cO_{(n)}\big(\cO'_{(0)}+\cO'_{(1)}\big)&=\big(\cO\cO'\big)_{(n+1)}+\big(\cO\cO'\big)_{(n)}\;,\\
\big(\cO'_{(0)}+\cO'_{(1)}\big)\cO_{(n)}&=\big(\cO'\cO\big)_{(n+1)}+\big(\cO'\cO\big)_{(n)}\;,    
\end{split}
\end{equation}
where the right-hand sides all belong to the space with $k>1$ that annihilates $\cH_1$. We can thus mod out the ideal and define the operator composition on the quotient space. In practice, given an operator \eqref{zerocharge} of charge zero, we do this by projecting out the components in the ideal:
\begin{equation}\label{restricted}
\cP(\cO):=\cO_{(0)}+\cO_{(1)} \;,
\end{equation}
which removes any term $\cO_{(k)}$ with $k>1$.
Projected operators obey
\begin{equation}\label{PP}
\cP(\cO)\cP(\cO')=\cP(\cO\cO')+\Big(\cP(\cO)\cP(\cO')\Big)_{(2)}\;,    
\end{equation}
as can be seen from \eqref{ideal}.
They do not form a subalgebra under composition but, since the extra term in \eqref{PP} belongs to the ideal, we can define the product on the quotient space by further projection:
\begin{equation}\label{comporestrict}
\cP\big(\cP(\cO)\cP(\cO')\big)=\cP\big(\cO\cO'\big)\;,  
\end{equation}
where we used \eqref{PP}.
This implies that operators \eqref{zerocharge} with $k\leq1$ form an associative algebra on $\cH_1$ upon using the projected product \eqref{comporestrict}. The vertex operators to be introduced in the following belong to this class and they will always be composed via the projected product \eqref{comporestrict}. To avoid cluttering the notation, the projected composition will be denoted by simple juxtaposition from now on.

\subsection{Yang-Mills vertex operators}

Upon color stripping, Yang-Mills theory exhibits a $C_\infty$ algebra structure \cite{Zeitlin:2008cc,Borsten:2021hua,Bonezzi:2022yuh}. This is a homotopy generalization of commutative associative algebras, consisting of a graded vector space $\cK$ endowed with a set of multilinear products $m_n:\cK^{\otimes n}\rightarrow\cK$. For the case of Yang-Mills, the graded vector space $\cK$ is isomorphic to the Hilbert subspace $\cH_1$, which we will review next. We will then show that all elements of $\cK$ can be generated by acting with vertex operators on a suitable vacuum state. Comparing to the general discussion of section \ref{sec:modules}, let us stress that the vertex operators to be discussed in the following \emph{cannot} be used directly to deform the BRST differential $Q$. This is related to the fact that the $C_\infty$ algebra on $\cK$ does not describe the field theory dynamics, prior to adding the color degrees of freedom. To deform $Q$, one has to tensor $\cK$ with the color Lie algebra $\mathfrak{g}$ and let the vertex operators also act as elements of ${\rm End}(\mathfrak{g})$, as done e.g.~in \cite{Zeitlin:2008cc,Bonezzi:2024emt}. 

Following the degree conventions of \cite{Bonezzi:2022bse}, the vector space $\cK$ is given by
\begin{equation}
\cK=\bigoplus_{i=0}^3\cK_i\;,\qquad\cK_i:=\cH_{1,i-1}\;,    
\end{equation}
which is a degree suspension of $\cH_1$. For every element $u\in\cK$ we define the differential $m_1$
by the action of the BRST operator:
\begin{equation}\label{m1}
m_1(u):=Q\,\ket{u}\;,    
\end{equation}
where we sometimes use the ket symbol to stress the action of operators on the Hilbert space $\cH_1$. The chain complex $(\cK,m_1)$ is displayed in the following diagram: 
\begin{equation}\label{K diagram}
\begin{tikzcd}[row sep=2mm]
\cK_{0}\arrow{r}{m_1}&\cK_{1}\arrow{r}{m_1}&\cK_{2}\arrow{r}{m_1}&\cK_{3}\\
\lambda&\cA&\cE&N\;,
\end{tikzcd}     
\end{equation}
with the elements in increasing degree corresponding to (color stripped) gauge parameters, fields, field equations and Noether identities, respectively. As vectors of $\cH_1$ they read
\begin{equation}\label{K states}
\begin{array}{ll}
\ket{\lambda}=\lambda(x)\,\cB\ket{0}\;, & \ket{\cA}=A_\mu(x)\,\alpha^\mu\ket{0}+\varphi(x)\,c\,\cB\ket{0}\;,\\[2mm]
\ket{\cE}=E_\mu(x)\,c\,\alpha^\mu\ket{0}+E(x)\,\cC\ket{0}\;,\quad&\ket{N}=N(x)\,c\,\cC\ket{0}\;.
\end{array}    
\end{equation}
This is the form of the Yang-Mills complex used in \cite{Bonezzi:2022yuh,Bonezzi:2022bse}, which includes an auxiliary scalar $\varphi$ together with its equation of motion $E$.
The action of $Q$ in \eqref{Q} gives the following expression for the differential:
\begin{equation}
\begin{split}
m_1(\lambda)&=\del_\mu\lambda\,\alpha^\mu\ket{0}+\B\lambda\,c\,\cB\ket{0}\;\in\cK_1\;,\\
m_1(\cA)&=(\B A_\mu-\del_\mu\varphi)\,c\,\alpha^\mu\ket{0}+(\del^\mu A_\mu-\varphi)\,\cC\ket{0}\;\in\cK_2\;,\\
m_1(\cE)&=(\B E-\del^\mu E_\mu)\,c\,\cC\ket{0}\;\in\cK_3\;,
\end{split}    
\end{equation}
which coincides with the one of \cite{Bonezzi:2022yuh}.

Having described the graded vector space and differential, in the following we will define a set of vertex operators corresponding to the states in $\cK$. This is useful to study the algebraic structures on $\cK$, since the space of operators ${\rm End}(\cK)$ is naturally equipped with a differential graded associative algebra. The associative product is just the composition of operators, while the differential is 
defined in terms of the graded 
commutator 
\begin{equation}
[A,B]:=AB-(-1)^{|A||B|}BA\;   
\end{equation}
and the BRST operator via $[Q,-]$. 
In order to define vertex operators, we first introduce a vacuum state in $\cK$ by acting with $\cB$ on the Fock vacuum:
\begin{equation}
\ket{1}:=\cB\ket{0}\;\in \; \cK_0\;.    
\end{equation}
Contrary to $\ket{0}\in\cH_0$, the state $\ket{1}$ does belong to $\cK$ and corresponds to a constant gauge parameter, obeying $Q\ket{1}=0$.
Given an arbitrary element $u\in\cK$, we want to find a vertex operator $V(u)\in{\rm End}(\cK)$, such that $u$ is generated by acting with $V(u)$ on the vacuum $\ket{1}$:
\begin{equation}\label{Vu}
V(u)\,\ket{1}=\ket{u} \quad \forall\;u\in\cK\;,\qquad |V(u)|=|u|\;,   
\end{equation}
where $|u|$ is the degree in $\cK$ as given in \eqref{K diagram}, while the degree $|V(u)|$ in the space of operators is given by ghost number.

Demanding the operator state correspondence \eqref{Vu} is not sufficient to determine $V(u)$, even restricting it to the space $l\leq1$ of \eqref{zerocharge}. While for any $V(u)$ obeying \eqref{Vu} one has $QV(u)\ket{1}=m_1(u)$, it is not true in general that $[Q,V(u)]=V(m_1(u))$, since they can differ by terms that annihilate $\ket{1}$. To fix the vertex operators we choose $V(\lambda)=\lambda$ as the starting point and demand that the chain map condition
\begin{equation}\label{ChainV}
[Q,V(u)]=V\big(m_1(u)\big)\quad\forall\;u\in\cK \;,   
\end{equation}
holds at the operator level. With these requirements we find
\begin{equation}\label{Vs}
\begin{split}
V(\lambda)&=\lambda(x)\;,\\
V(\cA)&=(\bar\cC\alpha^\mu+\cC\bar\alpha^\mu)A_\mu(x)+c\,\varphi(x)+2\,c\,\big(A^\mu(x)\del_\mu+f_{\mu\nu}(x)\,\alpha^\mu\bar\alpha^\nu\big)\;,\\
V(\cE)&=c\,(\bar\cC\alpha^\mu+\cC\bar\alpha^\mu)E_\mu(x)+\cC\bar\cC\,E(x)-2\,c\,\cC\bar\alpha^\mu\big(E_\mu(x)-\del_\mu E(x)\big)\;,\\
V(N)&=c\,\cC\bar\cC\,N(x)\;,
\end{split}
\end{equation}
where $f_{\mu\nu}=\del_\mu A_\nu-\del_\nu A_\mu$ is the abelian field strength. 
The vertex operators \eqref{Vs} can be viewed as the image of a map $V$ from $\cK$ to the space of operators:
\begin{equation}
V:\cK\rightarrow{\rm End}(\cK)\;,\quad |V|=0\;.    
\end{equation}
As a map, $V$ has intrinsic degree zero, given that $|V(u)|=|u|$. Using (the linear part of) the map differential $\del$ introduced in section \ref{sec:modules}, the condition \eqref{ChainV} is stated as the closure $\del V=0$, since evaluating $\del V:\cK\rightarrow{\rm End}(\cK)$ on $u$ gives
\begin{equation}\label{closedV}
\del V(u):= [Q,V(u)]-V\big(m_1(u)\big)=0\;.   
\end{equation}

\subsection{$C_\infty$ algebra from vertex operators}

In this section we will take advantage of the associative algebra of operators in ${\rm End}(\cK)$ to construct an $A_\infty$ algebra on the chain complex $(\cK,m_1)$. We will then show that the $C_\infty$ algebra of Yang-Mills theory, in the form presented in \cite{Bonezzi:2022yuh}, is obtained by a suitable shift of the products.

Using the correspondence \eqref{Vu}, we define a bilinear product $\mu_2:\cK\otimes\cK\rightarrow\cK$ by composing two vertex operators on the vacuum:
\begin{equation}\label{mu2}
\mu_2(u,v):=V(u)V(v)\ket{1}=V(u)\ket{v} \;.   
\end{equation}
The product $\mu_2$ has intrinsic degree zero and can be computed explicitly using the vertex operators in \eqref{Vs}, yielding
\begin{equation}\label{mu2 list}
\begin{split}
\mu_2(\lambda_1,\lambda_2)&=(\lambda_1\lambda_2)\,\cB\ket{0}\;,\\
\mu_2(\lambda,\cA)&=(\lambda A_\mu)\,\alpha^\mu\ket{0}+(\lambda\varphi)\,c\,\cB\ket{0}\;,\quad \mu_2(\cA,\lambda)=(\lambda A_\mu)\,\alpha^\mu\ket{0}+(\lambda\varphi+2\,A^\mu\del_\mu\lambda)\,c\,\cB\ket{0}\;,\\
\mu_2(\lambda,\cE)&=(\lambda E_\mu)\,c\,\alpha^\mu\ket{0}+(\lambda E)\,\cC\ket{0}\;,\quad \mu_2(\cE,\lambda)=\mu_2(\lambda,\cE)\;,\\
\mu_2(\cA_1,\cA_2)&=\big(\varphi_1 A_{2\mu}-\varphi_2 A_{1\mu}+2\,(A_1* A_2)_\mu\big)\,c\,\alpha^\mu\ket{0}+(A_1\cdot A_{2})\,\cC\ket{0}\;,\\
\mu_2(\cA,\cE)&=(-A^\mu E_\mu+\varphi E+2\,A^\mu\del_\mu E)\,c\,\cC\ket{0}\;,\quad \mu_2(\cE,\cA)=\mu_2(\cA,\cE)\;,\\
\mu_2(\lambda,N)&=(\lambda\,N)\,c\,\cC\ket{0}\;,\quad \mu_2(N,\lambda)=\mu_2(\lambda,N)\;,
\end{split}    
\end{equation}
where a dot denotes contraction with the metric: $A_1\cdot A_2=A^\mu_1 A_{\mu2}$, while the product $*$ between two vector fields is a generalization of the Lie bracket, which has the same formal structure as the generalized Lie derivative of double field theory \cite{Siegel:1993xq,Hull:2009mi,Hull:2009zb,Hohm:2010jy,Hohm:2010pp}:
\begin{equation}
(X*Y)^\mu:=X^\nu\del_\nu Y^\mu+(\del^\mu X_\nu-\del_\nu X^\mu)\, Y^\nu\;.    
\end{equation}
The definition \eqref{mu2} and chain map property \eqref{closedV} are enough to ensure that the differential $m_1$ is a derivation of $\mu_2$:
\begin{equation}\label{m1 Leibniz mu2}
\begin{split}
m_1\big(\mu_2(u,v)\big)&=Q V(u)V(v)\ket{1}=[Q, V(u)V(v)]\ket{1}\\
&=[Q, V(u)]V(v)\ket{1}+(-1)^{|u|} V(u)[Q,V(v)]\ket{1}\\
&=V\big(m_1(u)\big)V(v)\ket{1}+(-1)^{|u|} V(u)V\big(m_1(v)\big)\ket{1}\\
&=\mu_2\big(m_1(u),v\big)+(-1)^{|u|}\mu_2\big(u,m_1(v)\big)\;, 
\end{split}    
\end{equation}
where we used (\ref{m1}) in the first line. 

Although the operator composition is associative, the product $\mu_2$ in general is not. From the definition \eqref{mu2} one can infer that $V\big(\mu_2(u,v)\big)\ket{1}=V(u)V(v)\ket{1}$, but
$V\big(\mu_2(u,v)\big)$ and $V(u)V(v)$ may  differ by an operator that annihilates the vacuum, which we denote by
\begin{equation}\label{O2}
\cO_2(u,v):=V\big(\mu_2(u,v)\big)-V(u)V(v)\;.    
\end{equation}
Using the Leibniz rule \eqref{m1 Leibniz mu2} and the chain map property \eqref{closedV} one can prove that $\cO_2$ is closed in the sense that $\del\cO_2=0$, where the differential for a degree zero map $\cO_2:\cK\otimes\cK\rightarrow{\rm End}(\cK)$ is defined by
\begin{equation}
\del\cO_2(u,v):=[Q,\cO_2(u,v)]-\cO_2(m_1u,v)-(-1)^{|u|}\cO_2(u,m_1v)\;.    
\end{equation}
Apart from cohomological obstructions, this suggests that $\cO_2$ is exact, i.e. $\cO_2=\del V_2$, where
\begin{equation}
\del V_2(u,v):=[Q,V_2(u,v)]+V_2(m_1u,v)+(-1)^{|u|}V_2(u,m_1v)\;,    
\end{equation}
for a degree $-1$ map $V_2:\cK\otimes\cK\rightarrow{\rm End}(\cK)$. This is indeed the case and can be checked by direct computation of \eqref{O2}, yielding
\begin{equation}
\begin{split}
V_2(\cA_1,\cA_2)&=4\,(A_{\mu1}A_{\nu2})\,c\,\alpha^{[\mu}\bar\alpha^{\nu]} \;,\\
V_2(\cA,\cE)&=V_2(\cE,\cA)=2\,(EA_\mu)\,c\,\cC\bar\alpha^\mu\;,
\end{split}    
\end{equation}
as the only non-vanishing components of $V_2(u,v)$. We have thus established that the linear vertex operators $V(u)$ close under composition, modulo an exact term involving a bilinear vertex $V_2(u,v)$:
\begin{equation}\label{Vmu2}
V\big(\mu_2(u,v)\big)=V(u)V(v)+\del V_2(u,v)\;.    
\end{equation}
This is sufficient to prove that $\mu_2$ is associative up to homotopy, with a three-product $\mu_3$ that can be determined explicitly, but we will postpone the proof to the more relevant case of the $C_\infty$ product $m_2$.

As it is defined in \eqref{mu2}, the product $\mu_2$ has no definite symmetry and it does not appear to be immediately related to the $C_\infty$ product of Yang-Mills theory. Rather, $\mu_2$ is closely related to the Lian-Zuckerman product \cite{Lian:1992mn} for light modes of the string introduced by Zeitlin in \cite{Zeitlin:2009tj}. In order to find the relation with the $C_\infty$ algebra of Yang-Mills, we perform an exact shift on $\mu_2$, which defines a new product $m_2$ by
\begin{equation}\label{mu2 m2}
m_2(u,v):=\mu_2(u,v)+m_1(f_2(u,v))
+f_2(m_1(u),v)
+(-1)^{|u|}f_2(u,m_1(v))\;,    
\end{equation}
for a bilinear map $f_2$ of degree $-1$. Since the shift is exact, it is guaranteed that $m_1$ is a derivation of $m_2$ as well. For the same reason, if $\mu_2$ is associative up to homotopy, so is $m_2$. The most general local $f_2$ is given by the six parameter family
\begin{equation}
\begin{split}
f_2(\cA_1,\cA_2)&=k_1\,(A_1\cdot A_{2})\,c\,\cB\ket{0}\;,\\
f_2(\lambda,\cE)&=k_2\,(\lambda E)\,c\,\cB\ket{0}\;,\quad\quad\; f_2(\cE,\lambda)= k_3\,(\lambda E)\,c\,\cB\ket{0}\;,\\
f_2(\cA,\cE)&=k_4\,(EA_\mu)\,c\,\alpha^\mu\ket{0}\;,\quad f_2(\cE,\cA)=k_5\,(EA_\mu)\,c\,\alpha^\mu\ket{0}\;,\\
f_2(\cE_1,\cE_2)&=k_6\,(E_1E_2)\,c\,\cC\ket{0}\;.
\end{split}    
\end{equation}
Thanks to the relation \eqref{mu2 m2}, the vertex operator for the product $m_2$ has the same form \eqref{Vmu2} as the one for $\mu_2$, with different bilinear vertex operators given by
\begin{equation}\label{V2 mu2 V2 m2}
V_2^{(m_2)}(u,v)=V_2^{(\mu_2)}(u,v)+V\big(f_2(u,v)\big)\;.    
\end{equation}

The shift \eqref{mu2 m2} is an $A_\infty$ morphism. This means that if $(m_1,\mu_2)$ are the one- and two-products of an $A_\infty$ algebra, $(m_1,m_2)$ are also products of an $A_\infty$ algebra which, in particular, have no definite symmetry under the exchange of inputs.
However, choosing $k_1=1$ makes $m_2$ graded commutative for any choice of $k_2=k_3$, $k_4=k_5$ and $k_6$. This yields a three parameter family of graded symmetric products obeying
\begin{equation}
m_2(u,v)=(-1)^{|u||v|}m_2(v,u)\;,    
\end{equation}
and at the same time shows that $\mu_2$ is commutative up to homotopy, which is also the case for the Lian-Zuckerman product \cite{Lian:1992mn}. Upon further choosing $k_i=1$ for all parameters one recovers the Yang-Mills $C_\infty$ product in the form presented in \cite{Bonezzi:2022yuh}:
\begin{equation}\label{m2 list}
\begin{split}
m_2(\lambda_1,\lambda_2)&=(\lambda_1\lambda_2)\,\cB\ket{0}\;,\\
m_2(\lambda,\cA)&=(\lambda A_\mu)\,\alpha^\mu\ket{0}+\del_\nu(\lambda A^\nu)\,c\,\cB\ket{0}\;,\\
m_2(\lambda,\cE)&=\lambda\,(E_\mu-\del_\mu E)\,c\,\alpha^\mu\ket{0}\;,\\
m_2(\cA_1,\cA_2)&=(A_1\bullet A_2)_\mu\,c\,\alpha^\mu\ket{0}\;,\\
m_2(\cA,\cE)&=A^\mu(\del_\mu E- E_\mu)\,c\,\cC\ket{0}\;,\\
m_2(\lambda,N)&=(\lambda\,N)\,c\,\cC\ket{0}\;,
\end{split}    
\end{equation}
where the antisymmetric $\bullet$ product introduced in \cite{Bonezzi:2022yuh} reads
\begin{equation}
(X\bullet Y)^\mu=\del\cdot X\, Y^\mu-\del\cdot Y\, X^\mu+(X*Y)^\mu-(Y*X)^\mu\;.    
\end{equation}
The three-parameter family of $C_\infty$ products with arbitrary $k_2$, $k_4$ and $k_6$ is related to the Yang-Mills one by nonlinear redefinitions of parameters, fields and equations.
Using the relation \eqref{V2 mu2 V2 m2} we find the bilinear vertex operators associated to the product \eqref{m2 list}:
\begin{equation}\label{V2s for m2}
\begin{split}
V_2(\cA_1,\cA_2)&=(A_1\cdot A_2)\,c+4\,(A_{\mu1}A_{\nu2})\,c\,\alpha^{[\mu}\bar\alpha^{\nu]} \;,\\
V_2(\lambda,\cE)&=(\lambda E)\,c\;,\\
V_2(\cA,\cE)&=(EA_\mu)\,c\,(\cC\bar\alpha^\mu+\bar\cC\alpha^\mu)\;,\\
V_2(\cE_1,\cE_2)&=(E_1E_2)\,c\,\cC\bar\cC\;,
\end{split}    
\end{equation}
which thus obeys
\begin{equation}\label{Vm2}
V\big(m_2(u,v)\big)=V(u)V(v)+\del V_2(u,v)\;.    
\end{equation}
Notice that the term $\del V_2(u,v)$ above does not annihilate the vacuum, due to the shift \eqref{V2 mu2 V2 m2}, and thus contributes to $m_2(u,v)$.
In the following we are going to use these maps, together with the natural associative algebra of endomorphisms, to prove that the space $\cK$ with products $\{m_n\}$ is a $C_\infty$ algebra. 

In order to prove the $C_\infty$ relations, let us introduce a compact notation that allows us to work with maps alone, without reference to the inputs. To this end, let us denote by $\fm_n$ arbitrary $n$-linear maps in $\cK$ and by $\cV_n$ arbitrary $n$-linear vertex operator maps, i.e.
\begin{equation}
\fm_n:\cK^{\otimes n}\longrightarrow\cK \;,\qquad
\cV_n:\cK^{\otimes n}\longrightarrow{\rm End}(\cK)\;.   
\end{equation}
The maps $\fm_n$ can be composed with each other or with a vertex map, which we write as
\begin{equation}\label{composition}
\begin{split}
\fm_n\circ(\fm_{k_1}\otimes\cdots\otimes\fm_{k_n})&: \cK^{\otimes (k_1+\ldots+k_n)}\longrightarrow\cK\;,\\
\cV_n\circ(\fm_{k_1}\otimes\cdots\otimes\fm_{k_n})&: \cK^{\otimes (k_1+\ldots+k_n)}\longrightarrow{\rm End}(\cK)\;.    
\end{split}   
\end{equation}
We also define an associative product of two vertex maps, $\cV_k$ and $\cV_l$, as the vertex map $\cV_k\cdot\cV_l$ corresponding to the product of the two vertex operators:
\begin{equation}\label{ass compo}
\begin{split}
\big(\cV_k\cdot\cV_l\big)&:\cK^{\otimes(k+l)}\longrightarrow{\rm End}(\cK)\;,\\
\big(\cV_k\cdot\cV_l\big)(u_1,\ldots,u_{k+l})&:=(-1)^{|\cV_l|(|u_1|+\ldots+|u_k|)}\cV_k(u_1,\ldots,u_k
)\,\cV_l(u_{k+1},\ldots,u_{k+l})\;. 
\end{split}
\end{equation}

On the maps $\fm_n$ and $\cV_n$ we define a degree $+1$ differential $\del$ as
\begin{equation}\label{del defined}
\begin{split}
\del\fm_n&:\cK^{\otimes n}\longrightarrow\cK \;,\qquad
\del\cV_n:\cK^{\otimes n}\longrightarrow{\rm End}(\cK)\;,\\
\del\fm_n(u_1,\ldots,u_n)&:=m_1\fm_n(u_1,\ldots,u_n)-(-1)^{|\fm_n|}\fm_n(m_1u_1,\ldots,u_n)-\ldots\\ &-(-1)^{|\fm_n|+\sum_{i=1}^{n-1}|u_i|}\fm_n(u_1,\ldots,m_1u_n)
\;,\\
\del\cV_n(u_1,\ldots,u_n)&:=[Q,\cV_n(u_1,\ldots,u_n)]-(-1)^{|\cV_n|}\cV_n(m_1u_1,\ldots,u_n)-\ldots\\
&-(-1)^{|\cV_n|+\sum_{i=1}^{n-1}|u_i|}\cV_n(u_1,\ldots,m_1u_n)\;.
\end{split}    
\end{equation}
The nilpotence of $m_1$ and $Q$, together with the Jacobi identity of the graded commutator, ensure that $\del^2=0$. By construction, $\del$ acts as a derivation on the composition:
\begin{equation}
\begin{split}
\del\Big(\fm_n\circ(\fm_{k_1}\otimes\cdots\otimes\fm_{k_n})\Big)&=\del\fm_n\circ(\fm_{k_1}\otimes\cdots\otimes\fm_{k_n})+(-1)^{|\fm_n|}\fm_n\circ(\del\fm_{k_1}\otimes\cdots\otimes\fm_{k_n}) \;,\\
&+\ldots+(-1)^{|\fm_n|+\sum_{i=1}^{n-1}|\fm_{k_i}|}\fm_n\circ(\fm_{k_1}\otimes\cdots\otimes\del\fm_{k_n})\;,\\
\del\Big(\cV_n\circ(\fm_{k_1}\otimes\cdots\otimes\fm_{k_n})\Big)&=\del\cV_n\circ(\fm_{k_1}\otimes\cdots\otimes\fm_{k_n})+(-1)^{|\cV_n|}\cV_n\circ(\del\fm_{k_1}\otimes\cdots\otimes\fm_{k_n}) \;,\\
&+\ldots+(-1)^{|\cV_n|+\sum_{i=1}^{n-1}|\fm_{k_i}|}\cV_n\circ(\fm_{k_1}\otimes\cdots\otimes\del\fm_{k_n})\;.
\end{split}    
\end{equation}
Finally, $\del$ is also a derivation of the $\cdot$ product:
\begin{equation}
\del(\cV_k\cdot\cV_l)=\del\cV_k\cdot\cV_l+(-1)^{|\cV_k|}\cV_k\cdot\del\cV_l  \;,  
\end{equation}
thanks to the fact that the graded commutator $[Q,-]$ is a derivation of the operator product.

Using this notation, the vertex operator \eqref{Vm2} for $m_2$ can be written as
\begin{equation}\label{Vm2 map}
V\circ m_2=V\cdot V+\del V_2\;,    
\end{equation}
recalling that $|V|=0$. From this relation we start proving the $C_\infty$ identities. First of all, the condition that $m_1$ is a derivation of the product $m_2$ is expressed by $\del m_2=0$, as  can be seen from the definition \eqref{del defined}. To prove this, we compute its vertex operator map and use the above properties of the differential:
\begin{equation}
\begin{split}
V\circ(\del m_2)&=\del(V\circ m_2)\\
&=\del(V\cdot V+\del V_2)=\del V\cdot V+V\cdot\del V=0\;,
\end{split}    
\end{equation}
upon using the chain map condition $\del V=0$.
The vertex operator map \eqref{Vs} is injective, meaning that $V(u)=0$ implies $u=0$. We thus conclude that $\del m_2=0$, as required.

The next step in the $C_\infty$ relations is to check associativity of the two-product.  To this end, we define the associator of $m_2$
\begin{equation}\label{Ass}
{\rm Ass}(u,v,w):=m_2\big(m_2(u,v),w\big)-m_2\big(u,m_2(v,w)\big)  \;,  
\end{equation}
in terms of the map ${\rm Ass}:\cK^{\otimes3}\rightarrow\cK$ of degree zero. As before, we compute its vertex operator:
\begin{equation}
V\circ{\rm Ass}=V\circ\Big(m_2\circ(m_2\otimes1-1\otimes m_2)\Big)\;,        
\end{equation}
where $1$ denotes the identity in $\cK$. We now use associativity of the composition and \eqref{Vm2 map} to obtain
\begin{equation}
\begin{split}
V\circ{\rm Ass}&=\big(V\circ m_2\big)\circ(m_2\otimes1-1\otimes m_2)\\
&=\big(V\cdot V+\del V_2\big)\circ(m_2\otimes1-1\otimes m_2)\\
&=(V\circ m_2)\cdot V-V\cdot(V\circ m_2)+\del V_2\circ(m_2\otimes1-1\otimes m_2)\;,
\end{split}    
\end{equation}
where we used the definition \eqref{ass compo} of $\cV_k\cdot\cV_l$ to distribute the two factors in the composition. We proceed by inserting \eqref{Vm2 map} for $V\circ m_2$ and use associativity of $\cV_k\cdot\cV_l$, together with $\del m_2=0$ and the chain map condition $\del V=0$ to find
\begin{equation}\label{M3}
\begin{split}
V\circ{\rm Ass}&=\del V_2\cdot V-V\cdot\del V_2+\del V_2\circ(m_2\otimes1-1\otimes m_2)\\
&=\del\Big(V_2\cdot V-V\cdot V_2+ V_2\circ(m_2\otimes1-1\otimes m_2)\Big)\\
&=\del M_3\;,\\
M_3&:=V_2\cdot V-V\cdot V_2+ V_2\circ(m_2\otimes1-1\otimes m_2)\;,
\end{split}    
\end{equation}
where $M_3$ has degree $-1$ and maps $\cK^{\otimes3}$ to ${\rm End}(\cK)$. This is sufficient to prove that $m_2$ is homotopy associative. To see this, we define a three-product $m_3$ by
\begin{equation}\label{m3}
m_3(u,v,w):=M_3(u,v,w)\ket{1}\;.    
\end{equation}
The associator \eqref{Ass} then yields
\begin{equation}
\begin{split}
&{\rm Ass}(u,v,w)=V\big({\rm Ass}(u,v,w)\big)\ket{1}=\del M_3(u,v,w)\ket{1}\\
&=m_1m_3(u,v,w)+m_3(m_1u,v,w)+(-1)^{|u|}m_3(u,m_1v,w)+(-1)^{|u|+|v|}m_3(u,v,m_1w)\\
&=\del m_3(u,v,w)\;,
\end{split}    
\end{equation}
upon using the definition \eqref{del defined} for $\del M_3$. We have thus proven that $m_2$ is homotopy associative. From \eqref{M3} and \eqref{m3} it turns out that the only non-vanishing component of $m_3$ is given by
\begin{equation}\label{m3 real}
\begin{split}
m_3(\cA_1,\cA_2,\cA_3)&=V_2(\cA_1,\cA_2)V(\cA_3)\ket{1}+V(\cA_1)V_2(\cA_2,\cA_3)\ket{1}\\    
&=\big(A_1\cdot A_2A_{3\mu}+A_3\cdot A_2A_{1\mu}-2\,A_1\cdot A_3A_{2\mu}\big)\,c\,\alpha^\mu\ket{0} \;,   
\end{split}
\end{equation}
which has the correct symmetry for a $C_\infty$ three-product.

Given that ${\rm Ass}=\del m_3$, applying the vertex $V$ we conclude that $\del(V\circ m_3-M_3)=0$. Again, modulo cohomological obstructions we expect the relation $V\circ m_3=M_3+\del V_3$, for a degree $-2$ trilinear vertex $V_3$. Knowing the three-product \eqref{m3} and the vertex combination $M_3$ we find $V_3=0$ by direct computation. The vertex operator for $m_3$ is then given by 
\begin{equation}\label{Vm3}
V\circ m_3=V_2\cdot V-V\cdot V_2+ V_2\circ(m_2\otimes1-1\otimes m_2)\;,    
\end{equation}
so that the three-product is effectively derived from at most bilinear operations. This can now be used to prove that $m_2$ and $m_3$ are mutually compatible, with no need for a four-product, thus exhausting the $C_\infty$ algebra. The compatibility between $m_2$ and $m_3$ is expressed as
\begin{equation}
m_3\circ\big(m_2\otimes1\otimes1-1\otimes m_2\otimes1+1\otimes1\otimes m_2\big)=m_2\circ\big(m_3\otimes 1+1\otimes m_3\big)\;,    
\end{equation}
for vanishing $m_4$. To establish the above relation we compute the vertex operator of the left-hand side:
\begin{equation}
\begin{split}
&V\circ m_3\circ\big(m_2\otimes1\otimes1-1\otimes m_2\otimes1+1\otimes1\otimes m_2\big)\\
&=\big(V_2\cdot V-V\cdot V_2+ V_2\circ(m_2\otimes1-1\otimes m_2)\big)\circ\big(m_2\otimes1\otimes1-1\otimes m_2\otimes1+1\otimes1\otimes m_2\big)\\
&=\big(V_2\circ(m_2\otimes1-1\otimes m_2)\big)\cdot V+V\cdot\big(V_2\circ(m_2\otimes1-1\otimes m_2)\big)+V_2\circ\big({\rm Ass}\otimes 1+1\otimes{\rm Ass}\big)\\
&+V_2\cdot\big(V\circ m_2\big)-\big(V\circ m_2\big)\cdot V_2\;,
\end{split}    
\end{equation}
upon organizing the various compositions. We now use $V\circ m_2=V\cdot V+\del V_2$, ${\rm Ass}=\del m_3$ and \eqref{Vm3} to continue:
\begin{equation}
\begin{split}
&V\circ m_3\circ\big(m_2\otimes1\otimes1-1\otimes m_2\otimes1+1\otimes1\otimes m_2\big)\\
&=\big(V\circ m_3\big)\cdot V+V\cdot \big(V\circ m_3\big)+V_2\circ\big(\del m_3\otimes 1+1\otimes\del m_3\big)+V_2\cdot\del V_2-\del V_2\cdot V_2\\
&=\big(V\cdot V+\del V_2\big)\circ\big(m_3\otimes1+1\otimes m_3\big)-\del\big(V_2\cdot V_2+V_2\circ(m_3\otimes1+1\otimes m_3)\big)\\
&=V\circ m_2\circ\big(m_3\otimes1+1\otimes m_3\big)-\del\big(V_2\cdot V_2+V_2\circ(m_3\otimes1+1\otimes m_3)\big)\;.
\end{split}    
\end{equation}
In general, the $\del$-exact expression above would give the four product. However, given the bilinear vertices \eqref{V2s for m2} and the non-vanishing three-product \eqref{m3 real}, we have that all combinations $V_2\cdot V_2$, $V_2\circ(m_3\otimes1)$ and $V_2\circ(1\otimes m_3)$ vanish, thus proving that $m_4=0$. Furthermore, since the only non-vanishing component of $m_3$ maps three inputs in $\cK_1$ to $\cK_2$, all compositions of two $m_3$'s are trivially zero, which concludes the analysis of the $C_\infty$ relations.

\section{Towards the kinematic BV$^\B_\infty$ algebra of Yang-Mills}

The $C_\infty$ algebra described in the previous section accounts for the familiar consistency relations of perturbative gauge theories, upon color stripping. Yang-Mills theory, however, possesses a much larger hidden structure, termed BV$_\infty^\B$ algebra in \cite{Reiterer:2019dys}, that is the candidate for the kinematic algebra underlying off-shell double copy constructions \cite{Bonezzi:2022bse,Borsten:2022vtg,Bonezzi:2023pox}.

As we have discussed in the previous section, the $C_\infty$ algebra consists of the graded vector space $\cK$ together with the products $\{m_n\}$. The first ingredient beyond this structure is a second nilpotent differential, denoted by $b$, of opposite degree to $m_1$ and obeying
\begin{equation}\label{b def}
b^2=0\;,\quad m_1b+b\,m_1=\B\;,\quad |b|=-1\;, 
\end{equation}
where $\B$ is the wave operator. Given such an operator, one can define a kinematic bracket $b_2$ via its failure to obey the Leibniz rule with respect to the product $m_2$:
\begin{equation}
b_2(u,v):=b\,m_2(u,v)-m_2(bu,v)-(-1)^{|u|}m_2(u,bv)\;,    
\end{equation}
or, without inputs,
\begin{equation}
b_2=d_bm_2:=b\circ m_2-m_2\circ(b\otimes1+1\otimes b)\;,    
\end{equation}
where $d_b$ is defined analogously to $\del$ and obeys
\begin{equation}\label{dbdelbox}
d_b^2=0\;,\quad [\del,d_b]=[\B,-]\;.    
\end{equation}
The $b$ operator can be viewed as a generalization of the odd BV Laplacian, with $b_2$ being a derived bracket similar to the BV antibracket. The bracket $b_2$ is compatible with the product $m_2$ (in the graded Poisson sense) if
\begin{equation}\label{Pois}
{\rm Pois}(u,v_1,v_2):=b_2\big(u,m_2(v_1,v_2)\big)-m_2\big(b_2(u,v_1),v_2\big)-(-1)^{|v_1|(|u|+1)}m_2\big(v_1,b_2(u,v_2)\big)\;,    
\end{equation}
vanishes, in which case one can prove that $b_2$ is a graded Lie bracket obeying the Jacobi identity.
Apart from the cases of Chern-Simons theory \cite{Ben-Shahar:2021zww,Borsten:2022vtg,Bonezzi:2022bse,Bonezzi:2024dlv} and self-dual Yang-Mills theory in light-cone gauge \cite{Monteiro:2011pc,Bonezzi:2023pox}, the Poisson compatibility holds only up to homotopy and further terms proportional to the three-product $m_3$. This opens up the intricacy of the kinematic BV$_\infty^\B$ algebra, which grows quite dramatically with higher relations. In the following we start to address it from the point of view of vertex operators.

\subsection{Vertex operators and the kinematic bracket}

Coming back to the operators acting on the Hilbert space $\cH_1\simeq\cK$, the natural choice for the $b$-operator is the $b$ ghost itself. The defining relations \eqref{b def} are obeyed thanks to
\begin{equation}
Q\,b+b\,Q=\B\;,    
\end{equation}
which can be checked from the definition \eqref{Q} of $Q$ and $b^2=0$. The action of $b$ on $\cK$ can be read off from $b(u):=b\ket{u}$, yielding
\begin{equation}
b\ket{\cA}=\varphi\,\cB\ket{0}\;,\quad    
b\ket{\cE}=E_\mu\,\alpha^\mu\ket{0}\;,\quad
b\ket{N}=N\,\cC\ket{0}\;,
\end{equation}
which can be visualized diagrammatically with a degree shift:
\begin{equation}
\begin{tikzcd}[row sep=2mm]
\cK_{0}\arrow{r}{m_1}&\cK_{1}\arrow{r}{m_1}&\cK_{2}\arrow{r}{m_1}&\cK_{3}\\
\lambda&A_\mu&E&\\
&\arrow{ul}{b}\varphi&\arrow{ul}{b}E_\mu&\arrow{ul}{b}N
\end{tikzcd}     
\end{equation}
as in \cite{Bonezzi:2022bse}. From the relations  $\ket{u}=V(u)\ket{1}$ and $b\ket{1}=0$ one can infer that $[b,V(u)]\ket{1}=V(bu)\ket{1}$. This, however, does not hold as an operator relation and we are led to define the degree $-1$ vertex $V_{-1}:=d_bV$, acting as
\begin{equation}\label{V-1}
V_{-1}(u)=[b,V(u)]-V(bu) \;.   
\end{equation}
The vertex operators $V_{-1}(u)$ annihilate the vacuum $\ket{1}$ and are given by
\begin{equation}
\begin{split}
V_{-1}(\lambda)&=0\;,\\
V_{-1}(\cA)&=2\,A^\mu\del_\mu+2\,(\del_\mu A_\nu-\del_\nu A_\mu)\,\alpha^\mu\bar\alpha^\nu\;,\\
V_{-1}(\cE)&=2\,\cC\bar\alpha^\mu\big(\del_\mu E-E_\mu\big)-2\,c\,\big(E^\mu\del_\mu+(\del_\mu E_\nu-\del_\nu E_\mu)\,\alpha^\mu\bar\alpha^\nu\big)\;,\\
V_{-1}(N)&=-2\,c\,\cC\bar\alpha^\mu\del_\mu N\;,
\end{split}
\end{equation}
upon using the definition \eqref{V-1} with \eqref{Vs}. The vertex $V_{-1}$ measures the failure of $V$ to be a chain map for the differential $b$ and encodes most of the kinematic bracket $b_2$. To see this we derive the vertex operator for the bracket $b_2=d_bm_2$, using \eqref{Vm2 map} and \eqref{dbdelbox}:
\begin{equation}\label{Vb2}
\begin{split}
V\circ b_2&=V\circ(d_bm_2)=d_b(V\circ m_2)-(d_bV)\circ m_2\\
&=d_b(V\cdot V+\del V_2)-(d_bV)\circ m_2\\
&=V_{-1}\cdot V+V\cdot V_{-1}-V_{-1}\circ m_2+[\B,V_2]-\del(d_bV_2)\;.
\end{split}    
\end{equation}
In particular one can see that in the associative case, where $V_2=0$, the bracket $b_2$ is non-vanishing only if $V_{-1}\neq0$.
We turn next to the Poisson compatibility between the bracket and the product, in order to find its implications for the vertex operators. For simplicity we will assume that the underlying product $m_2$ is associative (as is the case, for instance, in Chern-Simons theory), so that $V_2=0$.

\subsection{Poisson compatibility}

The failure of the Poisson compatibility, parameterized by the ``Poissonator'' \eqref{Pois}, is the failure of the operator
\begin{equation}
b_u:=b_2(u,-)\;,    
\end{equation}
to be a derivation of the product $m_2$. This can be expressed as
\begin{equation}
{\rm Pois}(u,-,-)=[b_u,m_2]:=b_u\circ m_2-m_2\circ(b_u\otimes1+1\otimes b_u)\;.
\end{equation}
In the associative case $V_2=0$ and the composition of vertex operators $V$ closes on $m_2$: $V\big(m_2(u,v)\big)=V(u)V(v)$, which implies $m_2(u,v)=V(u)\ket{v}$. We further assume that $m_2$ is graded commutative, which is equivalent to $[V(u),V(v)]=0$. Applying \eqref{Vb2}, the vertex operator for the bracket $b_2$ reads
\begin{equation}\label{Vb2ass}
V\big(b_2(u,v)\big)=V_{-1}(u)V(v)+(-1)^{|u|}V(u)V_{-1}(v)-V_{-1}\big(m_2(u,v)\big)\;,    
\end{equation}
and, acting on the vacuum, one finds
\begin{equation}
b_2(u,v)=V_{-1}(u)V(v)\ket{1}=V_{-1}(u)\ket{v} \;,   
\end{equation}
recalling that $V_{-1}(u)\ket{1}=0$ for all $u$. This allows us to identify $b_u=b_2(u,-)$ with the vertex operator $V_{-1}(u)$. The Poissonator \eqref{Pois} can then be written as
\begin{equation}
\begin{split}
{\rm Pois}(u,v,w)&=b_um_2(v,w)-m_2(b_uv,w)-(-1)^{|v|(|u|+1)}m_2(v,b_uw)\\
&=V_{-1}(u)V(v)\ket{w}-V\big(b_uv\big)\ket{w}-(-1)^{|v|(|u|+1)}V(v)V_{-1}(u)\ket{w}\;.    
\end{split}
\end{equation}
This shows that the Poisson compatibility is equivalent to the condition \begin{equation}\label{PoisAssComp}
[V_{-1}(u),V(v)]-V\big(b_2(u,v)\big)=0\;,    
\end{equation}
on vertex operators. For the above relation to be consistent the product $m_2$ has to be graded commutative, since the differential $\del$ of the left-hand side of \eqref{PoisAssComp} is proportional to $[V(u),V(v)]$.
We will show below  how such a structure is realized for the toy model of Chern-Simons theory.

The simplest departure from the strict compatibility \eqref{PoisAssComp}
is to demand the vertex operators to instead obey
\begin{equation}
[V_{-1}(u),V(v)]-V\big(b_2(u,v)\big)=\del\Theta_2(u,v)  \;,  
\end{equation}
where $\Theta_2:\cK^{\otimes2}\rightarrow{\rm End}(\cK)$ is a degree $-2$ vertex map. If this is the case, the Poissonator \eqref{Pois} vanishes up to homotopy:
\begin{equation}
{\rm Pois}(u,v,w)=\del\theta_3(u,v,w)\;,\quad{\rm where}\quad \theta_3(u,v,w):=\Theta_2(u,v)\ket{w}\;.    
\end{equation}

\subsection{An example: Chern-Simons theory}

To exemplify the above discussion with a simple toy model, we consider the case of Chern-Simons theory in three dimensional flat space. In this case the graded vector space $\cK$ consists of functions $\omega(x,c)$ of spacetime coordinates $x^\mu$ and anticommuting odd coordinates $c^\mu$ of ghost number $+1$. At fixed ghost number $p=0,\ldots,3$ one has a $p$-form in target space:
\begin{equation}\label{omega p}
\ket{\omega_p}\equiv\omega_p(x,c)=\frac{1}{p!}\,\omega_{\mu_1\cdots\mu_p}(x)\,c^{\mu_1}\cdots c^{\mu_p}  \;,  
\end{equation}
and one identifies the (color-stripped) gauge field in ghost number $+1$. Similar to the Yang-Mills case discussed in the previous section, the graded functions $\omega(x,c)$ arise as the BRST Hilbert space of a topological worldline theory \cite{Witten:1992fb,Berkovits:2001rb}. 
The physical vacuum state is again identified as a constant gauge parameter in ghost number zero which, in the representation \eqref{omega p}, is just the constant function: $\ket{1}=1$. 

Operators acting on this Hilbert space are differential operators in $\del_\mu$ and $\frac{\del}{\del c^\mu}$, with coefficients depending on $x^\mu$ and $c^\mu$. The BRST operator $Q=c^\mu\del_\mu$ is just the de Rham differential and we write $Q\ket{\omega}=d\omega$.
Vertex operators, generating states out of the unit vacuum, are identified with the functions $\omega(x,c)$ themselves, since
\begin{equation}
\ket{\omega}=V(\omega)\ket{1}=\omega(x,c)\ket{1}=\omega(x,c) \;, 
\end{equation}
in a somewhat redundant notation. The vertex $V$ is a chain map for the differential, since
\begin{equation}
[Q,V(\omega)]=V(d\omega)\;.    
\end{equation}
Here it is useful to keep the distinction between $\omega\in\cK$ and $V(\omega)\in{\rm End}(\cK)$: although their explicit expression is the same, $V(\omega)=\omega(x,c)$ is thought of as the operator acting via pointwise multiplication of graded functions, which is why $[Q,-]$ is the appropriate differential.
The associative product is just the wedge product of forms: $m_2(\omega,\eta)=\omega\wedge\eta$, given by composition of vertex operators:
\begin{equation}
V(\omega\wedge\eta)=V(\omega)V(\eta)\;, 
\end{equation}
which fits with our requirement $V\circ m_2=V\cdot V$ for an associative product.

Although there is no elementary $b$-ghost, one can define a nilpotent $b$-operator via
\begin{equation}
b=\del^\mu\frac{\del}{\del c^\mu}\;,    
\end{equation}
which is the divergence operator on $p$-forms: $b\ket{\omega}=d^\dagger\omega$ and obeys $[Q,b]=\B$.
Using these definitions, the kinematic bracket $b_2$ is given by
\begin{equation}
\begin{split}
b_2(\omega,\eta)&=d^\dagger(\omega\wedge\eta)-(d^\dagger\omega)\wedge\eta-(-1)^{|\omega|}\omega\wedge d^\dagger\eta\\
&=\frac{\del\omega}{\del c_\mu}\,\del_\mu\eta+(-1)^{|\omega||\eta|}\frac{\del\eta}{\del c_\mu}\,\del_\mu\omega\;,
\end{split}   
\end{equation}
which coincides with the Schouten-Nijenhuis bracket of polyvectors \cite{Borsten:2022vtg,Bonezzi:2022bse,Bonezzi:2024dlv}, upon using the metric to identify forms and polyvector fields.
The vertex $V_{-1}(\omega)$ is the differential operator
\begin{equation}
\begin{split}
V_{-1}(\omega)&=[b,V(\omega)]-V(d^\dagger\omega)\\
&=\frac{\del\omega}{\del c_\mu}\,\del_\mu+(-1)^{|\omega|}\del_\mu\omega\,\frac{\del}{\del c_\mu}\;,     
\end{split}   
\end{equation}
and it is easy to see that it obeys the Poisson compatibility condition
\begin{equation}[V_{-1}(\omega),V(\eta)]-V\big(b_2(\omega,\eta)\big)=0\;.    
\end{equation}

\section{Conclusions}

In this paper we have taken first steps in order to derive the 
so-called kinematic algebra of Yang-Mills theory, which is a homotopy algebra 
with higher products,  from a simple associative algebra of operators (endomorphisms) 
on a larger vector space. We showed that the $C_{\infty}$ algebra carried by  the Yang-Mills 
kinematic space ${\cal K}$ can be injected, via vertex operators encoding $A_{\infty}$ morphisms, 
into the associative algebra of operators acting on the Hilbert space of a suitable worldline theory. 
Importantly for color-kinematics duality and the double copy construction of gravity from gauge theory, 
there is also a vast hidden algebraic structure on ${\cal K}$ called BV$^\B_{\infty}$, 
which is only partially understood. We took first steps in deriving also the associated maps 
from the vertex operators, but it is not clear yet whether this will eventually simplify  the derivation 
of these maps. Such progress would be needed in order to push the construction of the BV$^\B_{\infty}$ 
algebra, and hence the proof of color-kinematic duality, to all orders.

While we presented an explicit $A_{\infty}$ morphism that injects the kinematic homotopy algebra ${\cal K}$ of 
Yang-Mills theory into the associative algebra of operators on a Hilbert space 
 this map is not actually a  quasi-isomorphism.  
Such quasi-isomorphisms can instead be constructed via strictification, as we review in the appendices, 
but these constructions  do not really provide a practical simplification. 
It thus remains as an outstanding open problem to find a simple enough realization of 
explicit operators on a vector space whose algebra is quasi-isomorphic to the 
highly involved BV$^\B_{\infty}$ algebra underlying color-kinematics duality and double copy.

 \section*{Acknowledgments} 

We thank Michael Reiterer, Ivo Sachs, Anton Zeitlin and Barton Zwiebach for discussions and in particular Felipe D\'iaz-Jaramillo for early collaboration. R.B. would also like to thank Fiorenzo Bastianelli, Olindo Corradini, Filippo Fecit and Emanuele Latini for discussions and collaboration on related projects.

\noindent
This work is funded   by the European Research Council (ERC) under the European Union's Horizon 2020 research and innovation programme (grant agreement No 771862)
and by the Deutsche Forschungsgemeinschaft (DFG, German Research Foundation), ``Rethinking Quantum Field Theory", Projektnummer 417533893/GRK2575. The work of R.B. is funded by the Deutsche Forschungsgemeinschaft (DFG, German Research
Foundation) Projektnummer 524744955. The work of C.C. is funded by by the Deutsche Forschungsgemeinschaft (DFG, German Research
Foundation), "Homological Quantum Field Theory", Projektnummer 9710005691.

\appendix

\section{Bar construction for $A_\infty$ modules}

In section 2 we saw that $A_\infty$ modules can be described as solutions to a Maurer-Cartan equation. Here, we will give an alternative description to $A_\infty$ modules. It is a variant of the bar construction: An $A_\infty$ algebra can be described in terms of a differential graded coalgebra. An $A_\infty$ module is equivalent to a certain differential graded comodule. For more details on this, the reader can consult for example \cite{keller2001,keller2006}.

\begin{defn}
Let $(C,\Delta,\epsilon)$ be a unital differential graded coalgebra. A $C$-comodule is a graded vector space $X$, together with a degree zero linear map
\begin{equation}
R: X \longrightarrow C \otimes X
\end{equation}
such that
\begin{equation}
(\mathrm{id}_C \otimes R) \circ R = (\Delta \otimes \mathrm{id}_X) \circ R \, , \quad (\epsilon \otimes \mathrm{id}_X) \circ R = R \, .
\end{equation}
We write $(X,R)$ for the data of a  $C$-comodule. A morphism $f$ of $C$-comodules is a degree zero map
\begin{equation}
f: (X,R_X) \longrightarrow (Y,R_Y) \, ,
\end{equation}
such that 
\begin{equation}
(\mathrm{id}_C \otimes f) \circ R_X = R_Y \circ f \, .
\end{equation}
\end{defn}
The above definition has an obvious generalization to the differential graded setting:
\begin{defn}
Given a unital differential graded coalgebra $(C,\text d,\Delta,\epsilon)$ and a $C$-comodule $(X,R)$, we say that $Q_X$ is a comodule differential on $(X,R)$, if
\begin{equation}
R: (X,Q_X) \longrightarrow (C \otimes X, \text d + Q_X)
\end{equation}
is a chain map. In that case, we say that $(X,Q_X,R)$ is a differential graded $C$-comodule. A morphism of differential graded $C$-comodules $f: (X,Q_X,R_X) \rightarrow (Y,Q_Y,R_Y)$ is a morphism of $C$-comodules that is also a chain map.
\end{defn}
\begin{defn}
Given a graded coalgebra $(C, \Delta,\epsilon)$ and a graded vector space $Y$, we define the $C$-comodule $F^l_C(Y)$ via
\begin{equation}
F^l_C(Y) =  C \otimes Y \, , \quad R = \Delta \otimes \text{id}_Y\, .
\end{equation}
Given any other $C$-comodule $(X,R_X)$ and a linear map 
\begin{equation}
f: X \rightarrow Y \, ,
\end{equation}
there is a unique associated morphism of $C$-comodules
\begin{equation}
F^l_C(f): (X,R_X) \longrightarrow (F_C(Y),R)
\end{equation}
given by
\begin{equation}
F_C(f) = (\text{id}_C \otimes f) \circ R_X \, .
\end{equation}
$F_C(f)$ is uniquely determined by $f$ via $(\epsilon \otimes \mathrm{id}_X) \circ F_C(f) = f$.
\end{defn}
\begin{prop}
Let $(C,\text d, \Delta,\epsilon)$ be a differential graded coalgebra and $X$ a graded vector space. Any degree one linear map $l: F_C(X) \rightarrow X$ determines a comodule differential $F_C(q) = q + \mathrm{d} \otimes \mathrm{id}_X$. Conversely, any comodule differential $Q$ on $F_C(X)$ is determined by a degree one map $q: F_C(X) \rightarrow X$ via $q = (\epsilon \otimes \mathrm{id}_X) \circ Q$.
\end{prop}
The following statement can be found in \cite{keller2001}, p.~13.
\begin{thm}
Given an $A_\infty$ algebra $(A,\{m_n\}_{n \ge 1})$ and a vector space $X$, there is a one-to-one correspondence between $A_\infty$-module structures on $X$ and comodule differentials on
\begin{equation}
F^l_{\mathcal{B}(A)}(X) = T^c(sA) \otimes X \, ,
\end{equation}
where $\mathcal{B}(A)$ is the unital differential graded coalgebra obtained from the bar construction on $A$.
\end{thm}
We saw above that any comodule differential on $T^c(sA) \otimes X$ is determined by a linear map
\begin{equation}
\mathcal{V}: T^c(sA) \otimes X \rightarrow X \, .
\end{equation}
Comparing to \eqref{ModuleDef} the term $\mathcal{V}_0$ constant in $A$ is the differential $Q$, while the higher terms $\mathcal{V}_{i \ge 1}$ are as before. 

The theorem allows us to give a straightforward definition of $A_\infty$ linear maps. 
\begin{defn}
Given an $A_\infty$ algebra $(A,\{m_k\}_{k \ge 1})$ and $A$-modules $(X,Q_X)$ and $(Y,Q_Y)$, an $A$-linear map is a morphism of differential graded comodules
\begin{equation}
F^l_{\mathcal{B}(A)}(f): F^l_{\mathcal{B}(A)}(X) \longrightarrow F^l_{\mathcal{B}(A)}(Y) \, .
\end{equation} 
We write $\hom_A(X,Y)$ for the space of $A$-linear maps from $X$ to $Y$. Note that $F^l_{\mathcal{B}(A)}(f)$ is equivalent to a degree zero linear map
\begin{equation}
f: T^c(sA) \otimes X \longrightarrow Y \, .
\end{equation}
\end{defn} 
The above construction follows the convention used in the main part of the paper, namely that the $A_\infty$ algebra acts on the space $X$ \emph{from the left}. The above constructions can also be made for the case when an $A_\infty$ algebra acts on a space $X$ from the right. A right $A$-module $X$ can be described by following the above construction after reordering the inputs
\begin{equation}
T^c(sA) \otimes X \Rightarrow X \otimes T^c(sA) \, .
\end{equation}
In this way, any left $A$-module structure on $X$ determines also a right $A$-module structure, so in reality there is no difference between the two concepts. However, below we will still use the concept of a right $A$-module independently from that of a left $A$-module. The reason for that will be explained below. We write
\begin{equation}
    F^l_C(X) = C \otimes X \, , \qquad F^r_C(X) = X \otimes C
\end{equation}
to distinguish between left and right modules.

\section{A strictification of $A_\infty$ algebras}

In this section, we will elaborate on how the action of an $A_\infty$ algebra $(A,\{m_n\}_{n \ge 1},e)$ on itself gives rise to a strictification. This strictification was used in \cite{Amorim_2016} in order to construct a tensor product of $A_\infty$ algebras. In that reference it is mentioned that this strictification arises from the space of right $A$-linear maps on $A$. In this section, we want to give more details on that idea. We start by considering the case of a strictly associative algebra without differential. 

\begin{defn}
Suppose we have a unital associative algebra $(A,m_2,e)$, where $e \in A$ is the unit. A right $A$-module is a vector space $X$ together with a map
\begin{equation}
{\mathcal V}_1: X \otimes A \rightarrow X \, ,
\end{equation}
such that
\begin{equation}
\mathcal{V}_1(\mathcal{V}_1(x,a),b) = \mathcal{V}_1(x,m_2(a,b)) \, , \quad \mathcal{V}_1(x,e) = x \, .
\end{equation}
As usual, we write $\mathcal{V}_1(x,a) = xa$. A right $A$-linear map $f: (X,\mathcal{V}^X_1) \rightarrow (Y,\mathcal{V}^Y_1)$ is a linear map $f: X \rightarrow Y$, such that
\begin{equation}
f(xa) = f(x)a \, .
\end{equation}
We write $\hom_A(X,Y)$ for the space of all right $A$-linear maps from $X$ to $Y$. We further define $\text{End}_X(A) := \hom_A(X,X)$. $\text{End}_A(X)$ becomes an algebra under composition.
\end{defn}
Recall the definition of $\text{End}(X)$ being the space of maps on a vector space $X$ linear in the underlying field. If $X$ is also a right $A$-module, we have that $\text{End}_A(X) \subseteq \text{End}(X)$ as a subalgebra. Any unital associative algebra $(A,m_2,e)$ becomes a right $A$-module upon defining
\begin{equation}
\mathcal{V}_1(a,b) = m_2(a,b) \, .
\end{equation}
Therefore, we can also define the algebra $\text{End}_A(A)$. 
\begin{prop}
Given a unital associative algebra $(A,m_2,e)$, we have an isomorphism of algebras $A \cong \mathrm{End}_A(A)$. Explicitly, the maps
\begin{equation}
\begin{split}
A &\longrightarrow \mathrm{End}_A(A) \, , \\
a &\longmapsto l_a := m_2(a,-)
\end{split}
\end{equation}
and
\begin{equation}\label{EvalonE}
\begin{split}
\mathrm{End}_A(A) &\longrightarrow A \, , \\
f &\longmapsto f(e)
\end{split}
\end{equation}
are morphism of algebras and inverse to each other.
\end{prop}
We want to stress that the left multiplication $l_a = m_2(a,-)$ is right $A$-linear as a map of modules, since
\begin{equation}
l_a(bc) = a(bc) = (ab)c = l_a(b)c \, .
\end{equation}
This means that $A$ carries both a left and a right $A$-module structure and these structures are $A$-linear with respect to each other. This is the reason for the statement made in the end of last section. It is useful to consider the notions of left and right $A$-modules independently, even when, strictly speaking, a left $A$-module is the same as a right $A$-module. On $A$, when expressing the right-module structure $\mathcal{V}_1^R = \mathcal{V}_1$ in terms of a left module structure $\mathcal{V}^L_1$, we define
\begin{equation}
\mathcal{V}^L_1(a,b) = \mathcal{V}^R_1(b,a) = ba \, .
\end{equation}
While this defines a perfectly fine left module structure, it is not the same as the one given by left multiplication $l_a$, unless
\begin{equation}
ba = \mathcal{V}^L_1(a,b) = l_a(b) = ab \, ,
\end{equation}
i.e.~unless $A$ is commutative. In other words, we could say that $A$ has two, in general inequivalent, left-module structures given by $l_a$ and $\mathcal{V}^L_1(a,-)$, which agree only when $A$ is commutative. We will not do this, since it is arguably more natural to think of the two structures on $A$ as defining independent left and right-module structures.

The above statement can be generalized to $A_\infty$ algebras in the following sense: There is a quasi-isomorphism from an $A_\infty$ algebra $(A,\{m_n\}_{n \ge 1})$ to the differential graded algebra $\text{End}^\infty_A(A)$ of right $A$-linear maps from $A$ to itself. We will give the definition of $\text{End}^\infty_A(A)$ below.

Recall that in the previous section we defined $A_\infty$-linear maps between modules over $A_\infty$-algebras as morphisms of differential graded comodules 
\begin{equation}
F^r_{\mathcal{B}(A)}(f): F^r_{\mathcal{B}(A)}(X) \longrightarrow F^r_{\mathcal{B}(A)}(Y) \, .
\end{equation}
Since these are determined by degree zero linear maps $f: X \otimes \mathcal{B}(A) \rightarrow Y$, they trivially are comodule morphisms. The only non-trivial statement is that they are also chain maps. The obstruction for $F_{\mathcal{B}(A)}(f)$ to be a chain map is given by
\begin{equation}
\partial_{X,Y} F^r_{\mathcal{B}(A)}(f) = (\mathcal{V}^Y + \text{id}_{Y} \otimes M_A) \circ F^r_{\mathcal{B}(A)}(f) - F^r_{\mathcal{B}(A)}(f) \circ (\mathcal{V}^X + \text{id}_{X} \otimes M_A) \, ,
\end{equation}
where $\mathcal V^X$ and $\mathcal V^Y$ are the comodule differentials on $X$ and $Y$ and $M_A$ the coderivation on $T^c(sA)$. This motivates the following definition.
\begin{defn}
Let $(A,\{m_n\}_{n \ge 1})$ be an $A_\infty$ algebra and $X$ and $Y$ right $A$-modules. We define $(\hom^\infty_A(X,Y),\partial)$ to be the space of comodule morphisms
\begin{equation}
F^r_{\mathcal{B}(A)}(f): F^r_{\mathcal{B}(A)}(X) \longrightarrow F^r_{\mathcal{B}(A)}(Y)
\end{equation}
of any degree, with differential $\partial_{X,Y}$ as above (including a sign in the commutator, depending on the degree of $f$ in the usual way). Since comodule morphisms can be added, we have that $\hom^\infty_A(X,Y)$ is a linear space and so $(\hom^\infty_A(X,Y),\partial_{X,Y})$ is a differential graded vector space.
\end{defn}
\begin{cor}
There is a composition
\begin{equation}
\mu_{X,Y,Z}: \hom^\infty_A(Y,Z) \otimes  \hom^\infty_A(X,Y) \longrightarrow \hom^\infty_A(X,Z) \, ,
\end{equation}
which is a chain map as well as associative, in the sense that
\begin{equation}
\mu_{W,Y,Z} \circ (\mathrm{id}_{\hom^\infty_A(Y,Z)} \otimes \mu_{W,X,Y}) = \mu_{W,X,Z} \circ (\mu_{X,Y,Z} \otimes \mathrm{id}_{\hom^\infty_A(W,X)})
\end{equation}
for any four modules $W,X,Y,Z$.
\end{cor}
Since $F^r_{\mathcal{B}(A)}(f) \in \hom^\infty_A(X,Y)$ can be equivalently described by the map $f: X \otimes T^c(sA) \rightarrow Y$, the differential and composition can also be described in terms of an action on $f$. For the differential $\partial_{X,Y}$, we have
\begin{equation}
\begin{split}
\partial_{X,Y} f(x,a_1,\ldots, a_n) = \sum_{i = 0}^n\mathcal{V}^Y_{n-i}(f(x,a_1,\ldots, a_i),a_{i+1},\ldots, a_n) \\
- \sum_{i = 0}^n(-)^f f(\mathcal{V}^X_i(x,a_1,\ldots,a_i),a_{i+1},\ldots, a_n) \\
- \sum_{k = 1}^n\sum_{i = 0}^{n-k} (-)^{f + a_1 + \ldots + a_i} f(x,a_1,\ldots,a_{i},m_k(a_{i+1},\ldots, a_{i+k}), a_{i + k + 1},\ldots, a_n) \, ,
\end{split}
\end{equation}
where $\mathcal{V}^{X}$ (resp.~$\mathcal{V}^{Y}$) denote the right action of $A$ on $X$ (resp.~$Y$). The composition is given by
\begin{equation}
\mu_{X,Y,Z}(g,f)(x,a_1,\ldots,a_n) = \sum_{i = 0}^n g(f(x,a_1,\ldots,a_i),a_{i+1},\ldots,a_n) \, .
\end{equation}
\begin{defn}
Given a right $A$-module $X$ over an $A_\infty$ algebra $(A,\{m_n\}_{n \ge 1})$, we define $\text{End}^\infty_A(X) := \hom_A^\infty(X,X)$. This is a differential graded algebra with differential $\partial_{X,X}$ and product $\mu_{X,X,X}$.
\end{defn}
\begin{prop}
An $A_\infty$ algebra is a right $A$-module over itself. Explicitly, the comodule differential $\mathcal V$ on $F^r_{\mathcal{B}(A)}(A) = A \otimes T^c(s A)$ is induced by the $A_\infty$-algebra products
\begin{equation}\label{Rightaction}
\mathcal{V}(x,a_1,\ldots,a_n) = m_{n+1}(x,a_1,\ldots, a_n) \, ,
\end{equation}
where $x \in A$ and $a_1,\ldots, a_n \in sA$.
\end{prop}
We will now give the main statement of this section, namely that for any unital $A_\infty$ algebra $(A,\{m_n\}_{n \ge 1},e)$, there is a quasi-isomorphism $A \cong \text{End}^{\infty}_A(A)$, where $A$-linearity in $\text{End}^{\infty}_A(A)$ is defined with respect to the $A$-module structure given in \eqref{Rightaction}. We first need to define the notion of a unital $A_\infty$ algebra.
\begin{defn}
A unital $A_\infty$ algebra $(A,\{m_n\}_{n \ge 1},e)$ is an $A_\infty$ algebra $(A,\{m_n\}_{n \ge 1})$, together with a degree zero element $e \in A$, such that
\begin{equation}
m_2(a,e) = m_2(e,a) = a
\end{equation}
for all $a \in A$ and 
\begin{equation}
m_n(a_1,\ldots,a_i = e,\ldots,a_n) = 0
\end{equation}
for all $n \ne 2$ and all $i$.
\end{defn}
\begin{thm}[{\bf Strictification of $A_{\infty}$}]
Let $(A,\{m_n\}_{n \ge 1},e)$ be a unital $A_\infty$ algebra. There is a quasi-isomorphism
\begin{equation}
\begin{split}
l: (A,\{m_n\}_{n \ge 1},e) &\longrightarrow (\mathrm{End}_A^\infty(A),\partial,\mu,\mathrm{id}_A) \, , \\
(a_1,\ldots,a_k) &\longmapsto l_{a_1,\ldots,a_k}
\end{split}
\end{equation}
of unital $A_\infty$ algebras, where $\partial := \partial_{A,A}$ and $\mu := \mu_{A,A,A}$ in the notation above and
\begin{equation}\label{Higherleftmult}
l_{a_1,\cdots,a_k}(x,b_1,\ldots,b_l) = m_{k+l+1}(a_1,\cdots,a_k,x,b_1,\cdots,b_l)
\end{equation}
with $k \ge 1$ and $l \ge 0$. There is also the inverse quasi-isomorphism
\begin{equation}
\begin{split}
\text{End}_A^\infty(A) &\longrightarrow A \, , \\
f &\longmapsto f(e) \, .
\end{split}
\end{equation}
\end{thm}
The theorem was proven in the more general case of filtered $A_\infty$ algebras in \cite{Amorim_2016}. To clarify things, let us explain the quasi-isomorphisms given in the theorem in more detail. The morphism $l$ maps $k$ elements $a_1,\ldots,a_k$ to the element $l_{a_1,\ldots,a_k} \in \hom(A \otimes T^c(sA),A)$, where it acts as in \eqref{Higherleftmult}. This is the generalization of the standard left-multiplication $a \mapsto l_a$, although $l_{a_1,\ldots,a_k}$ acts with multiple elements at the same time. In particular, when acting on $A \subseteq A \otimes T^c(sA)$, we have that $l_{a_1,\ldots,a_k}(x)$ is the action of $A$ as a left-module on itself. The higher terms with $l_{a_1,\ldots,a_k}(x,b_1,\ldots,b_l)$ with $l \ge 1$ are homotopy corrections to this left-action.

The inverse quasi-isomorphism is particularly simple. It takes a in general multilinear map $f: A \otimes T^c(sA) \rightarrow A$ to $f(e)$. Further, there are no non-linear parts in this $A_\infty$-morphism, i.e. no terms taking multiple $(f_1,\ldots,f_k)$ as inputs. This is the generalization of the map $\eqref{EvalonE}$.

To close this section, we want to explain the relation of the case of an ordinary graded algebra $A$ given in the beginning. It can be shown that $H^0(\text{End}^\infty_A(A)) \cong \text{End}_A(A)$, while all other cohomologies vanish. The construction of $\hom^\infty_A(X,Y)$ arises when one wants to find the Ext-groups $\text{Ext}^n_A(X,Y)$ for given $A$-modules $X$ and $Y$. Explicitly, $\text{Ext}^n_A(X,Y) = H^n(\hom^\infty_A(X,Y))$. On homology, the induced compositions $\mu_{X,Y,Z}$ are known as Yoneda products. From this perspective, the fact that $H^0(\text{End}^\infty_A(A)) \cong \text{End}_A(A)$ and $H^{k \ge 1}(\text{End}^\infty_A(A)) = 0$ immediately follows from $A$ being free over itself as a right $A$-module. In homological algebra, the elements in the homology of $\hom^\infty_A(X,Y)$ are considered the correct morphisms from $X$ to $Y$, while $\hom_A(X,Y)$ are the more naive ones. The reasons for that go beyond the scope of this paper. The upshot is that in homological algebra it is better to consider $\hom^\infty_A(X,Y)$ rather than $\hom_A(X,Y)$, even when $A$, $X$ and $Y$ are not differential-graded. Only when $X$ is free as an $A$-module, $\hom^\infty_A(X,Y)$ reduces to $\hom_A(X,Y)$ on homology, as it is the case with $\text{End}_A^\infty(A)$ and $\text{End}_A(A)$. For more details on $\text{Ext}_A(X,Y)$ and the Yoneda product, the reader can consult for example \cite{MayNotes}.

As a takeaway massage for a theoretical physicist, it could prove advantageous to consider non-linear vertex operators, even when the algebras they describe are strict, at least for the cases where the space of states $X$ is not the algebra $A$ itself.

\bibliographystyle{utphys}

\end{document}